\documentclass[]{elsarticle}

\usepackage{lineno}
\usepackage[colorlinks=true]{hyperref}
\usepackage{color}
\usepackage{amssymb}
\usepackage{tcolorbox}
\usepackage[percent]{overpic}
\usepackage{adjustbox}
\usepackage{tikz}
\usetikzlibrary{shapes, arrows.meta, positioning}

\modulolinenumbers[1]

\journal{Physica A}

\newtheorem{rem}{Remark}

\newcommand{\mysize}{\footnotesize}

\newcommand{\dn}{\textsc n}
\newcommand{\df}{\textsc f}
\newcommand{\db}{\textsc b}
\newcommand{\da}{\textsc a}

\newcommand{\Rmin}{R_{\mysize\textsc {min}}}
\newcommand{\Rmax}{R_{\mysize\textsc {max}}}

\newcommand{\Sref}{S^{\mysize\textsc{ref}}}

\newcommand{\Dcomfort}{D^{\mysize\textsc{comf}}}
\newcommand{\Dcontact}{D^{\mysize\textsc{contact}}}

\newcommand{\Dpushing}{D^{\mysize\textsc{push}}}
\newcommand{\Dminimal}{D^{\mysize\textsc{min}}}

\begin{document}

\begin{frontmatter}

\title{An All-Densities Pedestrian Simulator Based on a Dynamic Evaluation of the Interpersonal Distances}

\author[mymainaddress]{E. Cristiani\corref{mycorrespondingauthor}}
\cortext[mycorrespondingauthor]{Corresponding author}
\ead{e.cristiani@iac.cnr.it}

\author[mymainaddress]{M. Menci}
\ead{m.menci@iac.cnr.it}

\author[mysecondaryaddress]{A. Malagnino}
\ead{a.malagnino@gae-engineering.com}

\author[mysecondaryaddress]{G. G. Amaro}
\ead{g.amaro@gae-engineering.com}

\address[mymainaddress]{Istituto per le Applicazioni del Calcolo, Consiglio Nazionale delle Ricerche, Via dei Taurini 19, 00185 Rome, Italy}
\address[mysecondaryaddress]{GAe Engineering S.r.l., Via Assietta 17, 10128 Turin, Italy}

\begin{abstract}
In this paper we deal with pedestrian modeling, aiming at simulating crowd behavior in normal and emergency scenarios, including highly congested mass events.
We are specifically concerned with a new agent-based, continuous-in-space, discrete-in-time, nondifferential model, where pedestrians have finite size and are compressible to a certain extent. The model also takes into account the pushing behavior appearing at extreme high densities. 
The main novelty is that pedestrians are not assumed to generate any kind of ``field'' in the space around which determines the behavior of the crowd.
Instead, the behavior of each pedestrian solely relies on its knowledge of the environment and the evaluation of interpersonal distances between it and the others. 
The model is able to reproduce the concave/concave fundamental diagram with a ``double hump'' (i.e.\ with a second peak) which shows up when body forces come into play.
We present several numerical tests (some of them being inspired by the recent ISO 20414 standard), which show how the model can reproduce classical self-organizing patterns.
\end{abstract}


\end{frontmatter}


\section{Introduction}

\paragraph{Context and motivations}
In this paper we deal with pedestrian modeling, aiming at simulating crowd behavior in normal ($\sim$0-3 ped/m$^2$), congested ($\sim$4-7 ped/m$^2$), and highly congested scenarios ($\sim$8-11 ped/m$^2$). 
Simulating highly congested mass events, including transition from normal to extreme conditions, is crucial because such events have been the scene of serious accidents in the past years \cite{reportUK}. Therefore crowd managers simulate the mass event \emph{before} it actually takes place to plan adequate safety and security measures.
Dedicated commercial software applications (crowd simulators) have still some limitations, especially when dealing with high densities. 
This is one of the reasons why research in the field of crowd understanding and crowd simulation is still very active. 

The paper explores the potential of a new, minimal, zeroth-order model, i.e.\ a model in which pedestrians cannot control neither their acceleration nor speed, but they can only decide to move (at a given, constant speed), stay or make small adjustments in the crowd. The model is aimed at showing a seamless transition from normal to highly congested scenarios, without the usage of a pre-calculated fundamental diagram. 

\paragraph{Relevant literature}
The study of crowds is a multidisciplinary area which have attracted the interest of mathematicians, physicists, engineers, and psychologists.
Crowd modeling started from the pioneering papers \cite{hirai1975proc, okazaki1979TAIJp1, henderson1974TR} in the '70s.
Since then, all types of models were proposed, spanning nanoscale, microscale (agent-based), mesoscale (kinetic), macroscale (fluid-dynamics), and multiscale, either differential (based on ordinary or partial differential equations) or nondifferential (discrete choice, cellular automata, lattice gas), either discrete or continuous in time and space.
Also, models can be first-order (i.e.\ velocity based) or second-order (i.e.\ acceleration based), with local or nonlocal interactions, with metric or topological interactions. 
People can be assumed point-like or having finite size. In the last case they can be circles, ellipses or a combination of the two, and \textit{ad hoc} contact-avoidance procedures are defined. 
Beside small-scale collision-avoidance maneuvers (local navigation), models are also distinguished in terms of global path planning, i.e.\ how pedestrians choose their path to reach their target, also depending on the degree of knowledge of the environment, prediction capabilities, visibility conditions, and occlusions.
A number of review papers  
\cite{aghamohammadi2020TRB, 
	bellomo2011SR,
	bellomo2022M3AS,
	chen2018TR,
	corbetta2023AR,
	dong2020TITS,
	duives2013TRC,
	eftimie2018chapter, 
	li2019PhA,
	martinezgil2017CS,
	papadimitriou2009TRF,
	yang2020GM}, 
meta-review papers
	\cite{haghani2020SSp1, 
	haghani2020SSp2, 
	haghani2021PhA}
and books 
	\cite{cristianibook, 
	rosinibook, 
	kachroobook, 
	maurybook} 
are now available,  
we refer the interested reader to these references for an introduction to the field. 
It is also useful to mention that models for pedestrians often stem from, and share features with those developed in the context of vehicular traffic \cite{rosinibook, helbing2001RMP}.

\medskip

More specifically, in this paper we are concerned with agent-based, continuous-in-space, discrete-in-time, nondifferential models where pedestrians have finite size and are compressible to a certain extent.
Therefore, we are in the framework of \emph{optimal step models} \cite{seitz2012PRE, dietrich2014JCS, seitz2015PhA, vonsivers2015TRB} and close to \emph{velocity-based models} \cite{papadimitriou2009TRF, paris2007EUROGRAPHICS, tang2016PhA} as well as \emph{discrete-choice models} \cite{antonini2006TRB, robin2009TRB}. 
In this kind of models, even if a floor field is possibly employed for long-range navigation, pedestrians are not assumed to be passively advected by a force field generated by the environment (target, walls, obstacles) and by pedestrians themselves, as it happens in the classical \emph{social force models} \cite{chen2018TR}; rather than that, they autonomously make decisions and interact with the environment to satisfy their design objectives \cite{martinezgil2017CS}. 
To this end, \emph{cognitive heuristics} are used to define the behavior of the agents \cite{seitz2016JRSI}.
In our opinion this approach is preferable since agents directly translate their needs and goals in personal actions, and simulation artifacts are more easily avoided.  

\medskip

We also consider the possibility that the crowd reaches \emph{high densities}. 
On this regard, the analysis of the literature is complicated by the fact that the related nomenclature is not well established: keywords as \emph{irrationality}, \emph{panic}, \emph{pushing behavior} are largely used and their underlying assumptions are often treated as common knowledge, but a deeper analysis shows the high level of ambiguity of these terms, as well as the overly simplistic nature of their assumptions \cite{haghani2019JAT}.

The experimental literature about dense crowds is quite scarce. 
The reason for that is twofold: first, crowd congestion is highly dangerous and it is difficult to perform controlled experiments in such situations; see a recent example in \cite{jin2021JAT}. 
Second, it is quite difficult to take accurate measurements of flux, velocity and density during the events.
Our main source of experimental information are the papers \cite{helbing2007PRE, johansson2008ACS} in which the authors were able to compute the fundamental diagram (i.e.\ the relationship between flux and density) until 10 ped/m$^2$. 
Interestingly, the fundamental diagram shows a concave/concave shape with a ``double hump'' (or a ``second peak''), meaning that the flux reaches two local maximum points. 
The second peak appears because people are so densely packed that they are moved involuntarily by the crowd (i.e.\ individual motion is replaced by mass motion) \cite{helbing2007PRE}.
See also \cite{jin2019TRC, lohner2017CD} for an experimental study that confirms a counter-intuitive increase of velocity at high densities.

Regarding mathematical models, instead, the literature is richer. 
Let us first mention the macroscopic approach proposed in \cite{colombo2005M2AS} (see also \cite{chalons2007SISC}) where the mathematical properties of a fundamental diagram with double hump are investigated.
Velocity-based models with pushing forces are presented in \cite{kim2013SIGGRAPH, kim2015TVC} in the context of computer graphics, while social force models with pushing forces are presented in \cite{alrashed2020CD, helbing2000N, helbing2002PED, moussaid2011PNAS, yu2007PRE}. 
The paper \cite{moussaid2011PNAS} also adds cognitive heuristics.
A macroscopic model with crowd pressure force based on the Hughes's model is proposed in \cite{liang2021TRB}. 
A multiscale model for contact avoidance in high densities was proposed in \cite{narain2009SIGGRAPH}, while a multiscale model based on smoothed particle hydrodynamic technique was proposed in \cite{vantoll2021CeG}. 
It also includes pushing forces and, notably, it is able to propagate material waves in the crowd due to the pushing behavior.

\medskip

Finally, let us mention that in 2020 it was published the International Standard ISO 20414 \cite{ISO20414}.
The document addresses the procedures for verification and validation
of evacuation models in the context of building fires and indicates 30 numerical tests to be performed.
This kind of analysis goes beyond the scope of the present paper, still we have considered some of the proposed tests in Section \ref{sec:numerics}.


%
%
\section{The model}\label{sec:themodel}

\subsection{Basic concepts}\label{sec:basicideas}
In this paper we propose an agent-based, continuous-in-space, discrete-in-time, nondifferential model with contact-avoidance features and pushing behavior. 
Similarly to cellular automata, it is a zeroth-order model: at each time step, pedestrians decide either to stay still or to make a step of fixed length in a certain direction.

Dynamics are described by simple cognitive heuristics.  
In this regard, we follow the paradigm described in \cite{seitz2017RGP} which ``argues that human decision making has to be based on evolutionary-developed cognitive capacities, such as the ability to estimate distances or predict movement based on previous movement cues. 
Furthermore, it suggests that humans do not make decisions based on mathematical optimization, but rather employ simple heuristics, which may or may not lead to the optimal solution and do not require unbounded computational power.''

The main novelty is that the model is solely based on the evaluation of \emph{interpersonal distances} among agents. 
In particular, pedestrians are described by means of two state variables: their position (as usual), and the minimum distance they can accept to keep from the people in front of them. This distance can vary in time and reflects psychological aspects of the movement in a crowd.
In addition, we assume that pedestrians start pushing when the interpersonal distance is reduced beyond a certain threshold. In this case, some energy can be transferred from one body to another.

Finally, and most important, the crowd does not generate any kind of ``field'' in the surrounding space. This means that agent is the only subject of the decisions, and decisions are made only on the basis of the knowledge of the environment, the target, and of what the agent can see and understand \textit{from its point of view}. 

\subsection{Cognitive heuristics}\label{sec:cognitiveheuristics}
The proposed algorithm, which will be presented in detail later in this Section, translates the following cognitive heuristics:
\begin{itemize}
\item people know the environment they are moving in, and have a target to reach in minimal time. They are able to compute the fastest path to the target from any point of the space and their head is always oriented towards the target along the fastest path. 
This uniquely defines a direction of motion and then what it is `ahead' and `behind' each pedestrian;
\item each agent has a minimum distance which accepts to keep between itself and the other agents ahead. 
Agents do not want to get closer than that distance from people ahead, so they stop when this limit is reached. 
This constraint provides for a collision-avoidance feature (like the repulsion force does in the social force model);
\item on the other hand, agents do not want to be overtaken by people behind them, in order not to increase their time to target. 
To this end, agents accept to modify their minimum accepted distance: for example, if a pedestrian $P$ is too closely approached by someone from behind, the $P$'s accepted distance is decreased.
As a consequence, $P$ gets closer to the people ahead, thus reducing the possibilities to be overtaken;
\item if not pushed, pedestrians are allowed to make a step along the fastest path to the target, unless this step brings them closer than the accepted minimum distance to some other pedestrian in front, or into some obstacle/wall;
\item if pushed, two cases are possible: 
\begin{enumerate}
\item if they have some space in front, they actually move ahead in the pushing direction (regardless of the target position). According to the experimental paper \cite{wang2018JSM}, we assume that the strength of the pushing force is proportional to the distance between the pushing and the pushed agents;
\item instead, if pushed pedestrians have not space in front of them, they simply try to maximize the distance from the nearest neighbor, in order to find the most comfortable position. 
If there is no space to move at all, they stay still.
\end{enumerate}
\end{itemize}

\begin{rem}
The pushing forces completely change the dynamics since they transmit themselves along the crowd as the domino effect, letting people accelerate in situations in which they have no intention to do that. 
This will be the key point to reproduce the fundamental diagram with double hump, see Test 3 in Section \ref{sec:T2}.
\end{rem}

\subsection{Setting, body shape, and time discretization}
We consider a crowd of $N>1$ pedestrians moving in a bounded domain $\Omega\subset\mathbb R^2$.
We assume that each agent is represented by a \emph{circle}, and that all agents are equal (these assumptions can be easily relaxed, see Section \ref{sec:generalizations}).
We assume that agents are compressible to a certain extent, so that their radius $R$ is variable in the interval $[\Rmin,\Rmax]$, with $\Rmax>\Rmin>0$, cfr.\ \cite{song2019IEEEA}. 
More precisely, $\Rmax$ represents the first distance at which contacts are perceived, while $\Rmin$ represents the smallest possible distance, beyond which the person begins suffocating.

We define a final time $T>0$ for the simulation and a small time step $\Delta t>0$.
Note that $\Delta t$ is actually a parameter of the model, meaning that it must not be chosen ``as small as possible'' as one should do in the numerical approximation of differential models. 
Moreover, it is certainly related to the \textit{reaction time} of humans, but it does not coincide with it. We think that a better interpretation of $\Delta t$ is the minimum time needed to interrupt an action (i.e., a step forward) once it has began.

We enumerate the time steps by $n=1,\ldots,n_T$, with $n_T:=\frac{T}{\Delta t}$ and we denote the position of the $k$-th agent at time step $n$ by $\mathbf X^n_k\in\mathbb R^2$.

Pedestrians are updated \textit{one after the other}, from $k=1$ to $k=N$. 
During computation, we always use the most up-to-date position of agents. 
This means that, when updating the position of agent $k$, we consider the just updated positions $\{\mathbf X_h^{n+1}\}$ for $h=1,\ldots,k-1$ and the not yet updated positions $\{\mathbf X_h^{n}\}$ for $h=k+1,\ldots,N$.

\subsection{Target}
We assume that the crowd has a common target to reach in minimal time. 
Given the domain $\Omega$, an obstacle set $\mathcal O\subset\Omega$, and a target set $\mathcal T\subset\Omega$, it is possible to compute the fastest route to the target from any point of the domain. The most common way to do it is not considering the presence of the crowd, i.e.\ each pedestrian computes the fastest route as if it was the only person in the domain.

In a convex domain without obstacles, the optimal path is a straight line joining the starting point and the closest point of the target. Otherwise the solution is less trivial and can be found by solving the \emph{eikonal equation}
\cite{cacace2014SISC, falconebook, sethianbook}.
	\begin{equation}\label{ee}
	\left\{
	\begin{array}{ll}
	\|\nabla u(\mathbf x)\|=1,& \mathbf x\in\Omega\backslash(\mathcal T\cup\mathcal O) \\
	u(\mathbf x)=+\infty, & \mathbf x\in\mathcal O\\
	u(\mathbf x)=0, & \mathbf x\in\mathcal T
	\end{array}
	\right. 
	\end{equation}
	and then moving in the opposite direction to the gradient of the solution
	$$
	\mathbf e(\mathbf x):=-\frac{\nabla u(\mathbf x)}{\|\nabla u(\mathbf x)\|}, \qquad \mathbf x\in\Omega\backslash(\mathcal T\cup\mathcal O).
	$$
	
	\medskip
	
	In Test 6, Section \ref{sec:T5}, we will also follow the approach originally proposed by Hughes in a macroscopic framework \cite{hughes2002TRB} (see also \cite[Sect.\ 4.4]{cristianibook} and \cite{hartmann2014proc}), modifying equation (\ref{ee}) as
	\begin{equation}\label{ee-modif}
  	\left\{
  \begin{array}{ll}
  s(\rho(\mathbf x))\|\nabla u(\mathbf x)\|=1,& \mathbf x\in\Omega\backslash(\mathcal T\cup\mathcal O) \\
  u(\mathbf x)=+\infty, & \mathbf x\in\mathcal O\\
  u(\mathbf x)=0, & \mathbf x\in\mathcal T
  \end{array}
  \right. 
	\end{equation}
	where $\rho$ is the density of the crowd at some fixed time and $s$ is a decreasing function which relates the speed of motion to the density of pedestrians. The idea is that the crowded regions slow down the motion, thus the fastest path must be computed taking into account also this fact. 
	In a microscopic framework the function $\rho$ can be computed by suitable convolution from the positions of the single agents.
	
	\medskip
	
	 For the sake of completeness, let us also mention that the fastest path to the target can be computed by pedestrians also considering the prediction they can do about the motion of the rest of the crowd, i.e.\ taking into account the position that the crowd will occupy during the whole journey. 
	 In this case, all the trajectories, at any time, are dependent of each other and then we enter the field of mean field games, see, e.g., \cite{lachapelle2011TRB, cristiani2023CMS}. 

\medskip

In the rest of the paper we will assume that a pedestrian located at a generic position $\mathbf x\in\mathbb R^2$ is oriented in the direction $\mathbf e(\mathbf x)$ and it looks in that direction. 
This orientation divides the space in two parts, the \emph{front space} (ahead)
$$
\mathcal F(\mathbf x):=\{\mathbf y\in\Omega \ : \ (\mathbf y-\mathbf x)\cdot \mathbf e(\mathbf x)\geq 0\}
$$
and the \emph{back space} (behind)
$$
\mathcal B(\mathbf x):=\{\mathbf y\in\Omega \ : \ (\mathbf y-\mathbf x)\cdot \mathbf e(\mathbf x)<0\}.
$$

\subsection{State variables and parameters}
In the proposed model, agents interact with each other on the basis of \emph{interpersonal distances} only. 
For any agent $k$, the following four distances are defined: 
\begin{itemize}
	\item $\dn_k(\mathbf x;\mathbf X)$ is the minimum distance between the agent $k$, assumed to be located at a generic point $\mathbf x\in\mathbb R^2$, and all the other $N-1$ agents, assumed to be located at $\mathbf X:=(\mathbf X_1,\ldots,\mathbf X_{k-1},\mathbf X_{k+1},\ldots,\mathbf X_N)\in\mathbb R^2\times\mathbb R^{N-1}$, 
	$$\dn_k(\mathbf x;\mathbf X):=\min\{\|\mathbf x-\mathbf X_h\|, h\neq k\}.$$ 
	\item $\df_k(\mathbf x;\mathbf X)$ is the minimum distance between the agent $k$, assumed to be located at a generic point $\mathbf x\in\mathbb R^2$, and all the other agents ahead of it (i.e.\ in $\mathcal F(\mathbf x)$),  
	$$
	\df_k(\mathbf x;\mathbf X):=\min\{\|\mathbf x-\mathbf X_h\|, \ \mathbf X_h\in\mathcal F(\mathbf x), \ h\neq k\}.
	$$
	Let us also denote by $f^k$ the (index of the) nearest neighbor ahead of (the agent indexed by) $k$. See Fig.\ \ref{fig:basics}.
	\item $\db_k(\mathbf x;\mathbf X)$ is the minimum distance between the agent $k$, assumed to be located at a generic point $\mathbf x\in\mathbb R^2$, and all the other agents behind it (i.e.\ in $\mathcal B(\mathbf x)$),  
	$$
	\db_k(\mathbf x;\mathbf X):=\min\{\|\mathbf x-\mathbf X_h\|, \ \mathbf X_h\in\mathcal B(\mathbf x), \ h\neq k\}.
	$$
	Similarly to before, let us also denote by $b^k$ the nearest neighbor behind $k$. See Fig.\ \ref{fig:basics}.
	\item $\da_k^n$ is instead the minimum distance the agent $k$ accepts to keep from the nearest neighbor in front of it at the time step $n$, and it is updated at every time step with a specific rule which will be described later on. 
	This distance is very important since it ultimately defines the behavior of the agent.  
\end{itemize}
\begin{figure}[h!]
	\centering
	\begin{overpic}[width=0.6\textwidth]{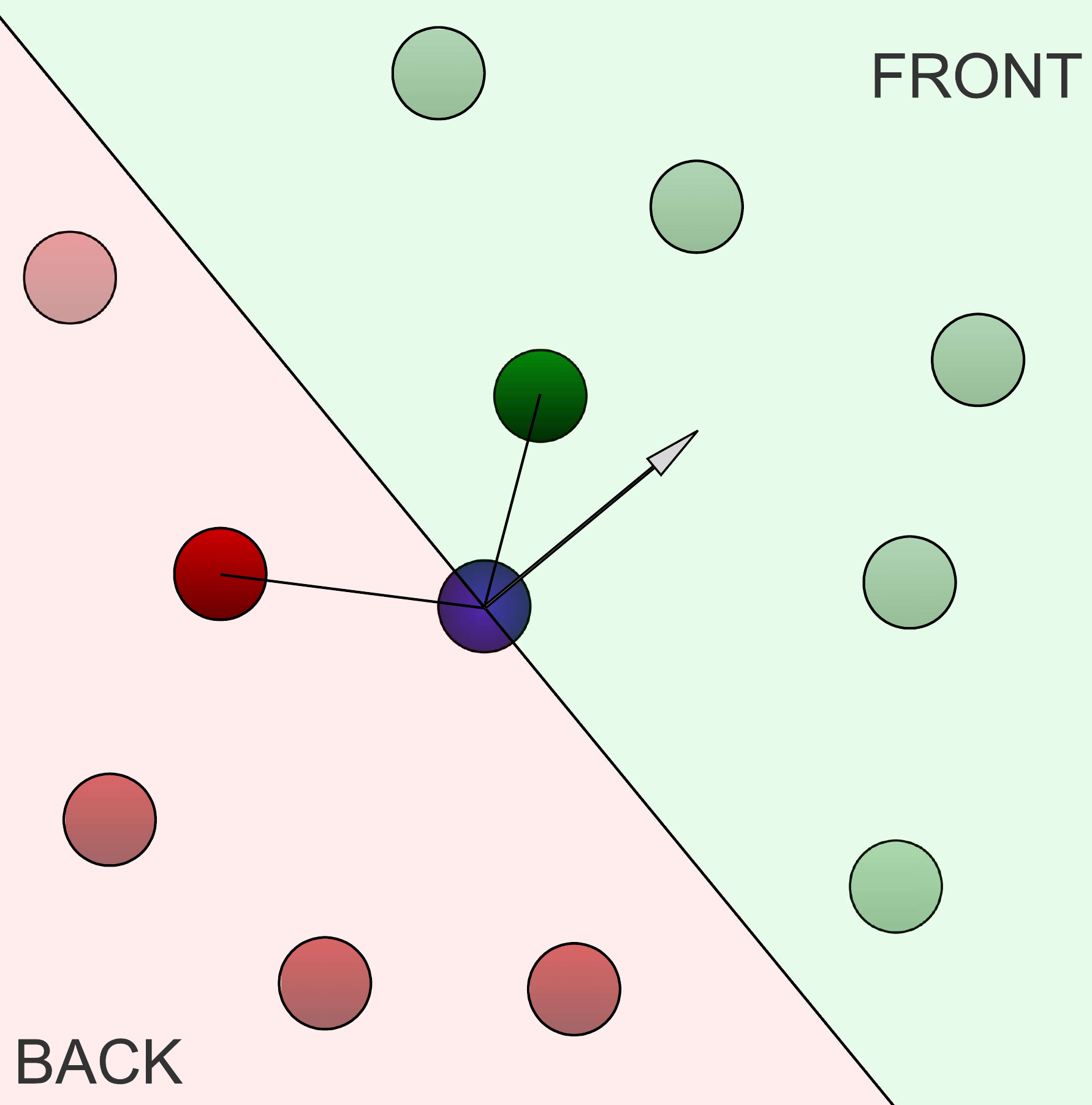}
		\put(41,36){$\mathbf x$} 
		\put(55,50){$\mathbf e(\mathbf x)$} 
		\put(5,12){$\mathcal B(\mathbf x)$} 
		\put(80,80){$\mathcal F(\mathbf x)$} 
		\put(41,54){$\df_k$}
		\put(40,63){$f^k$}
		\put(30,43){$\db_k$}
		\put(11,44){$b^k$}
	\end{overpic}
	\caption{agent $k$ in $\mathbf x$, its orientation $\mathbf e(\mathbf x)$ and its nearest neighbors ahead and behind.}
	\label{fig:basics}
\end{figure}
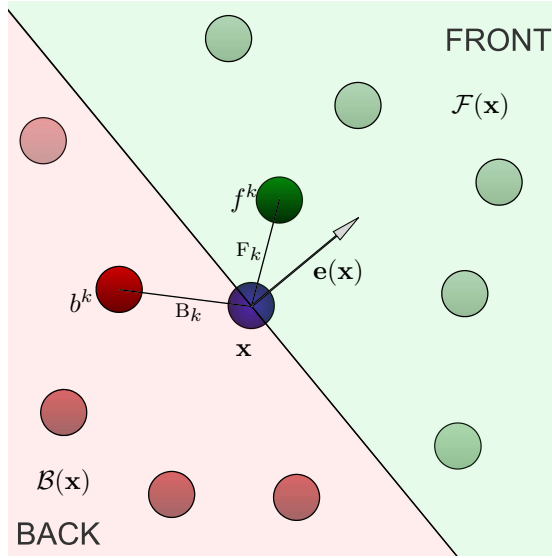
\begin{rem} 
It is useful to maintain the theoretical possibility that the agent $k$ at time step $n$ is located in a temporary/tentative position $\mathbf x\neq\mathbf X^n_k$. This is the reason why we have defined $\dn$, $\df$, and $\db$ as a function of a generic position $\mathbf x$. 
\end{rem}

\medskip

The following four parameters $\Dcomfort >\Dcontact >\Dpushing> \Dminimal$ are also defined: 
\begin{itemize}
	\item The \textit{comfort distance} $\Dcomfort$ is the minimum distance the agent $k$ would like to keep, ideally, from the nearest neighbor. 
	\item The \textit{contact distance} $\Dcontact:=2\Rmax$ is the distance at which it is touched by another agent.
	\item The \textit{pushing distance} $\Dpushing$ is the distance at which an agent starts to be pushed by the neighbor behind it.
	\item The \textit{minimal distance} $\Dminimal:=2\Rmin$ is the minimal distance the agent $k$ can have from each other. Note that this is only a \textit{soft constraint}, i.e.\ the algorithm temporarily allows a shorter distance among the agents. This is important to give the right amount of elasticity to the system.
\end{itemize} 

Finally, we define four additional parameters:
\begin{itemize}
	\item $\Sref$ is the reference speed. At any time step, the agents can either stay still or move with this speed. (Lower speeds can be only achieved, on average on multiple time steps, alternating standstill with motion.)
	\item $\Delta x:=\Sref \Delta t$ is the displacement in one time step in normal conditions.
	\item $0<\varepsilon\leq 1$ is the reduction factor for the displacement in high density conditions, with respect to normal conditions. 
	\item $\alpha>0$ is a parameter which translates the attention the pedestrians pay not to be overtaken by people behind.  
	\item $C$ tunes the strength of the pushing force.
\end{itemize}

\subsection{The algorithm}\label{sec:thealgorithm}
At the initial time, the positions of all agents are defined, and the accepted distance of every agent $k$ is initialized with $\da_k^1=\Dcomfort$. 
For any $n=1,\ldots n_T-1$ and $k=1,\ldots,N$, let us define
$$\mathbf X_{(k)}^n:=(\mathbf X_{1}^{n+1},\ldots,\mathbf X_{k-1}^{n+1},\mathbf X_{k+1}^{n},\ldots,\mathbf X_{N}^{n}),$$
then the algorithm follows the flowchart in Fig.\ \ref{fig:diagrammadiflusso}, with the following blocks:
 \begin{figure}[h!]
 	\begin{center}
 \begin{tikzpicture}[font=\small,thick]
 %
\node[draw,
minimum width=2cm,
minimum height=0.5cm
] (block1) {\begin{tabular}{c}
 \texttt{if} $\db_k(\mathbf X_k^n)\leq \alpha\df_k(\mathbf X_k^n)$ \texttt{then} $\da_k^{n+1}=\db_k(\mathbf X_k^n)$ \texttt{else} $\da_k^{n+1}=\da_k^n$.  \\ [2mm]
 \texttt{project} $\da_k^{n+1} \texttt{ onto } [\Dcontact,\Dcomfort]$ \\ [2mm]
 \end{tabular}
};
%
\node[draw,
diamond,
below=of block1,
minimum width=2.5cm,
inner sep=0] (block2) {
	$\db_k(\mathbf X_k^n)\geq\Dpushing$
};
\node[draw,
fill=blue!5!white,
below left=of block2,
minimum width=4.0cm,
minimum height=0.7cm,
inner sep=0] (block3) {
	\texttt{compute} $X_k^*$ \texttt{with block 1}
};
\node[draw,
diamond,
below right=of block2,
minimum width=2.5cm,
inner sep=0] (block4) {
	$\df_k(\mathbf X_k^n)>  \Dminimal$ 
};
\node[draw,
fill=green!5!white,
below right=of block4,
minimum width=4cm,
minimum height=0.7cm,
inner sep=0] (block5) {
	\texttt{compute} $X_k^*$ \texttt{with block 3}
};
\node[draw,
fill=red!5!white,
left=of block5,
minimum width=4cm,
minimum height=0.7cm,
inner sep=0] (block6) {
	\texttt{compute} $X_k^*$ \texttt{with block 2}
};
\node[draw,
below=2cm of block6,
minimum width=2cm,
minimum height=0.7cm,
inner sep=0] (block7) {\texttt{set} $X_k^{n+1}=X_k^*$};
 \draw[-latex] (block1) edge (block2);
 \draw[-latex] (block2) -| (block3) node[pos=0.75,fill=white,inner sep=2]{yes};
 \draw[-latex] (block2) -| (block4) node[pos=0.75,fill=white,inner sep=2]{no};
 \draw[-latex] (block4) -| (block6) node[pos=0.75,fill=white,inner sep=2]{yes};
\draw[-latex] (block4) -| (block5) node[pos=0.75,fill=white,inner sep=2]{no};
 \draw[-latex] (block3) edge node[]{}(block7);
 \draw[-latex] (block5) edge node[]{}(block7);
 \draw[-latex] (block6) edge node[]{}(block7);
 \end{tikzpicture} 
\caption{Flowchart of the algorithm (time step $n$, agent $k$).}
\label{fig:diagrammadiflusso}
\end{center}
\end{figure}
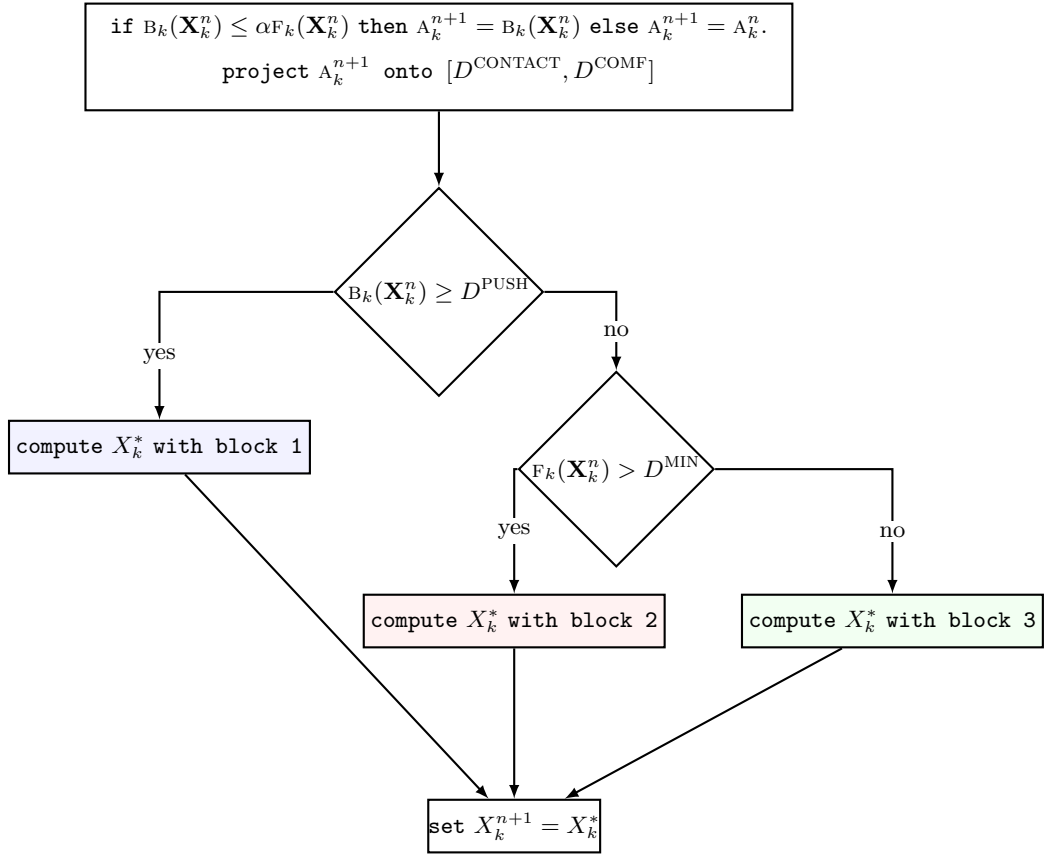

\begin{tcolorbox}[colback=blue!5!white]
	\texttt{BLOCK 1 (NORMAL STEP)} \\
	Among the points of the set 
	$$\{\mathbf X^n_k\}\cup\{\mathbf X^n_k+\Delta x(\cos \theta, \sin \theta) : \theta\in[0,2\pi)\}$$ 
	find the point $\mathbf X^*_k$ such that $u(\mathbf X^*_k)$ is minimal,  under the constraints
	$$\mathbf X^*_k\in\Omega\backslash\mathcal O  \ \textrm{and} \ \df_k(\mathbf X_k^*;\mathbf X_{(k)}^n)\geq\da^{n+1}_k.$$ 
\end{tcolorbox}

\begin{tcolorbox}[colback=red!5!white]
	\texttt{BLOCK 2 (PUSHING)} \\
	Set $$\mathbf X^*_k:=\mathbf X^n_k+
	\left\{
	\begin{array}{ll}
	C\Delta t (\mathbf X^n_k-\mathbf X^{n+1}_{b^k}), & \textrm{if } b^k< k \\ [2mm]
	C\Delta t  (\mathbf X^n_k-\mathbf X^{n}_{b^k}), & \textrm{if } b^k> k.
	\end{array}
	\right.	
	$$ 
\end{tcolorbox}

\begin{tcolorbox}[colback=green!5!white]
	\texttt{BLOCK 3 (FIND SPACE)} \\
	Among the points of the set 
	$$\{\mathbf X^n_k\}\cup\{\mathbf X^n_k+\varepsilon\Delta x(\cos \theta, \sin \theta) : \theta\in[0,2\pi)\}$$ 
	find the point $\mathbf X^*_k$ such that $\mathbf X^*_k\in\Omega\backslash\mathcal O$ and $\dn_k(\mathbf X^*_k;\mathbf X_{(k)}^n)$ is maximal. 
\end{tcolorbox}


\subsection*{Some details}
At the beginning, the accepted distance $\da_k$ is updated. If someone is approaching from behind ($\db_k\leq \alpha\df_k$), $\da_k$ takes the value of $\db_k$, otherwise it remains unchanged. 
The new value of $\da_k$ is then projected in the interval $[\Dcontact,\Dcomfort]$, meaning that if it is lower than $\Dcontact$ it becomes equal to $\Dcontact$, if instead it is larger than $\Dcomfort$ it becomes equal to $\Dcomfort$.

\begin{figure}[h!]
	\centering
	\begin{overpic}[width=0.7\textwidth]{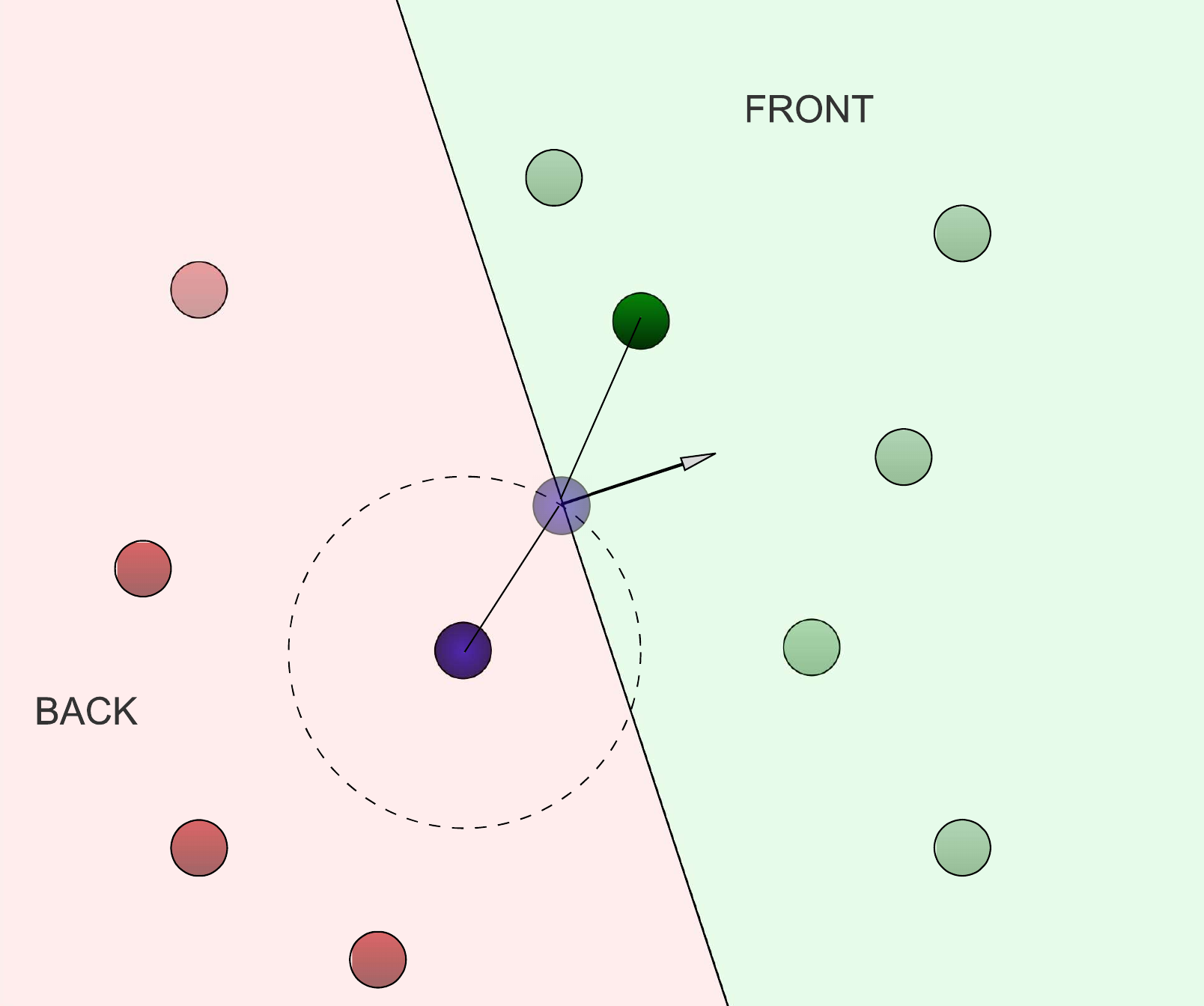}
		\put(37,23){$\mathbf X_k^n$} \put(45,36){$\mathbf x$}
		\put(45,50){$\df_k$}
		\put(50,60){$f^k$}
		\put(36,35){$\Delta x$}
  		\put(52,47){$\mathbf e(\mathbf x)$}
  		\put(3,15){$\mathcal B(\mathbf x)$} 
  		\put(75,73){$\mathcal F(\mathbf x)$} 
\end{overpic}
	\caption{\texttt{BLOCK 1} (normal step): agent $k$ tries to move at distance $\Delta x$ from its current position $\mathbf X_k^n$ (dark blue circle). 
	In the tentative position (light blue circle) $\mathbf x$ it evaluates $\mathbf e(\mathbf x)$, $\mathcal F(\mathbf x)$ and then it finds the nearest neighbor ahead $f^k$ and the distance $\df_k$ from it.
	It is important to temporary remove $k$ from $\mathbf X_k^n$ and put it in $\mathbf x$, otherwise one risks that agent $k$ finds itself as nearest neighbor ahead of it!}
	\label{fig:block1a}
\end{figure}

\begin{figure}[h!]
	\centering
	\begin{overpic}[width=0.7\textwidth]{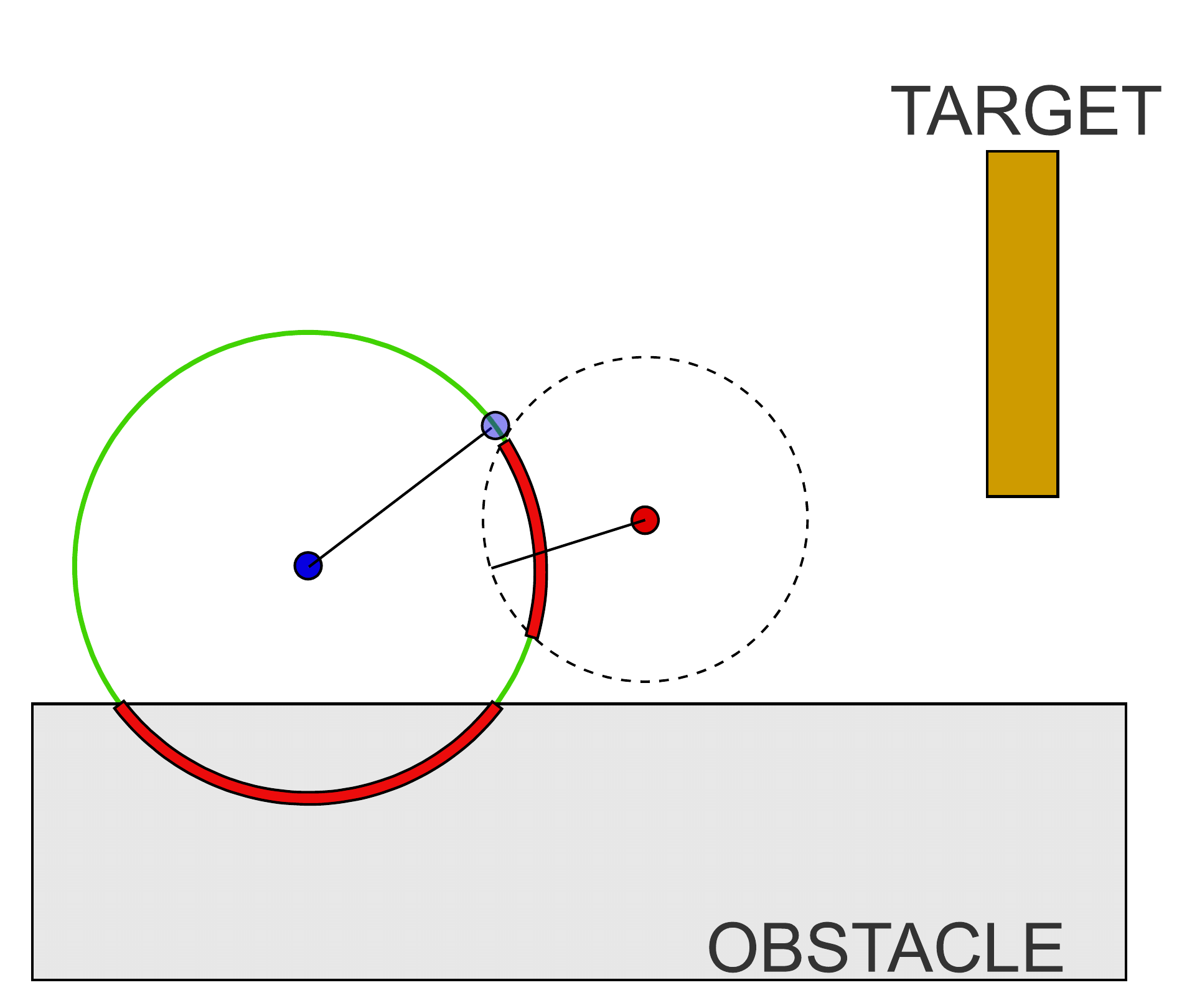}
		\put(21,32){$\mathbf X_k^n$} \put(39,52){$\mathbf X_k^*$}
		\put(56,40){$f^k$}
		\put(47,41){$\df_k$}
		\put(29,44){$\Delta x$}
	\end{overpic}
	\caption{\texttt{BLOCK 1} (normal step): agent $k$ have found its next position $\mathbf X_k^{n+1}=\mathbf X_k^*$ (light blue circle). It is not inside the obstacle, it is sufficiently far form the nearest neighbor ahead $f^k$, and it minimizes the distance to target. 
	Red portions of the circle are prohibited positions. 
	}
	\label{fig:block1b}
\end{figure}
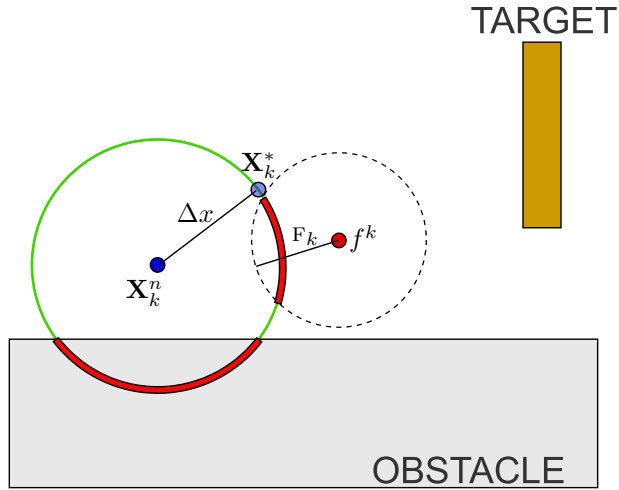

\begin{figure}[h!]
	\centering
	\begin{tabular}{cc}
	 \begin{overpic}[width=0.45\textwidth]{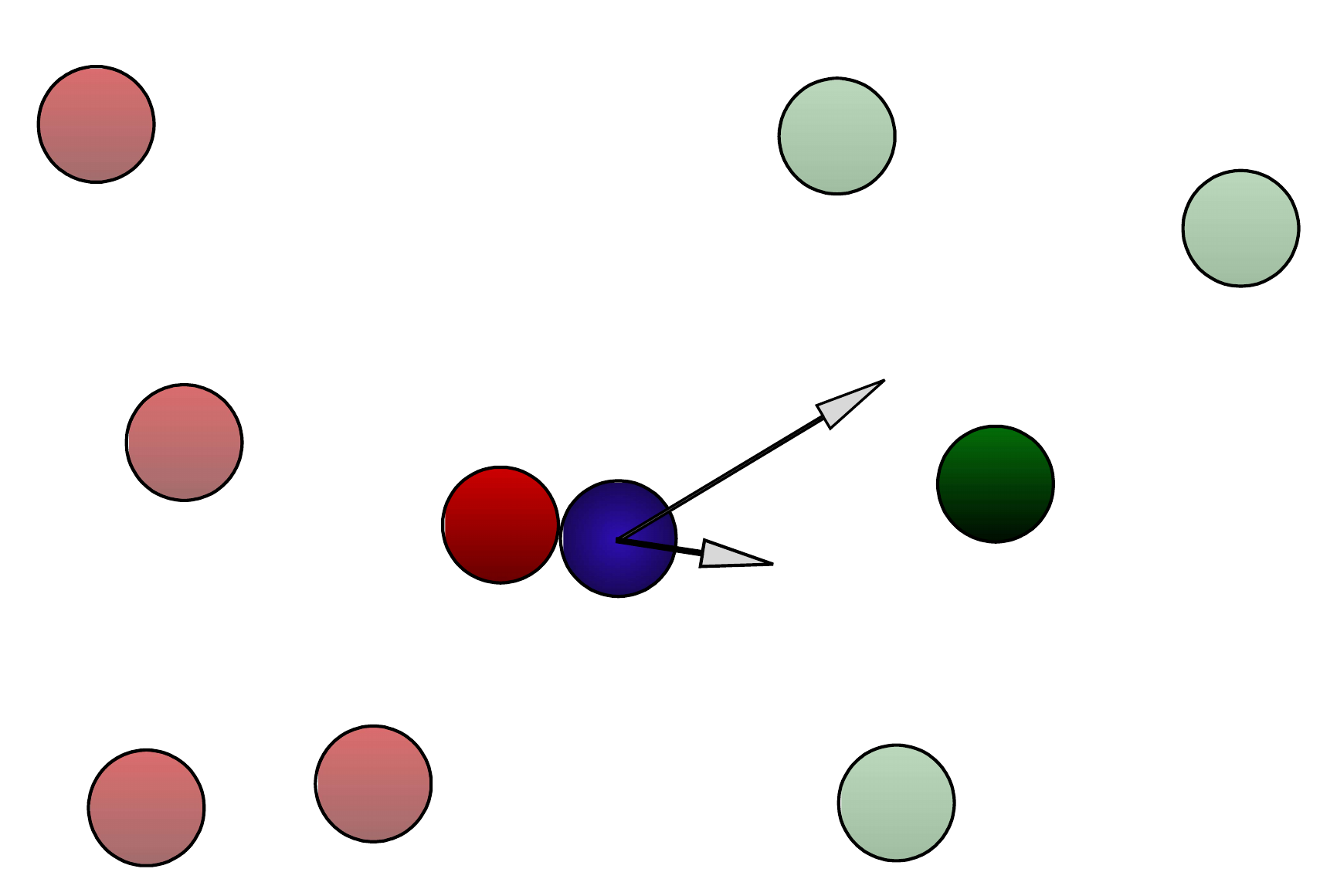}
		\put(12,45){$\mathcal B(\mathbf X_k^n)$} 
		\put(80,17){$\mathcal F(\mathbf X_k^n)$} 
		\put(48,40){$\mathbf e(\mathbf X_k^n)$} 
		\put(60,24){$\mathbf p$} 
		\put(80,33){$f^k$}
		\put(27,22){$b^k$}
		\put(44,16){$\mathbf X_k$}
	\end{overpic} &
\begin{overpic}[width=0.45\textwidth]{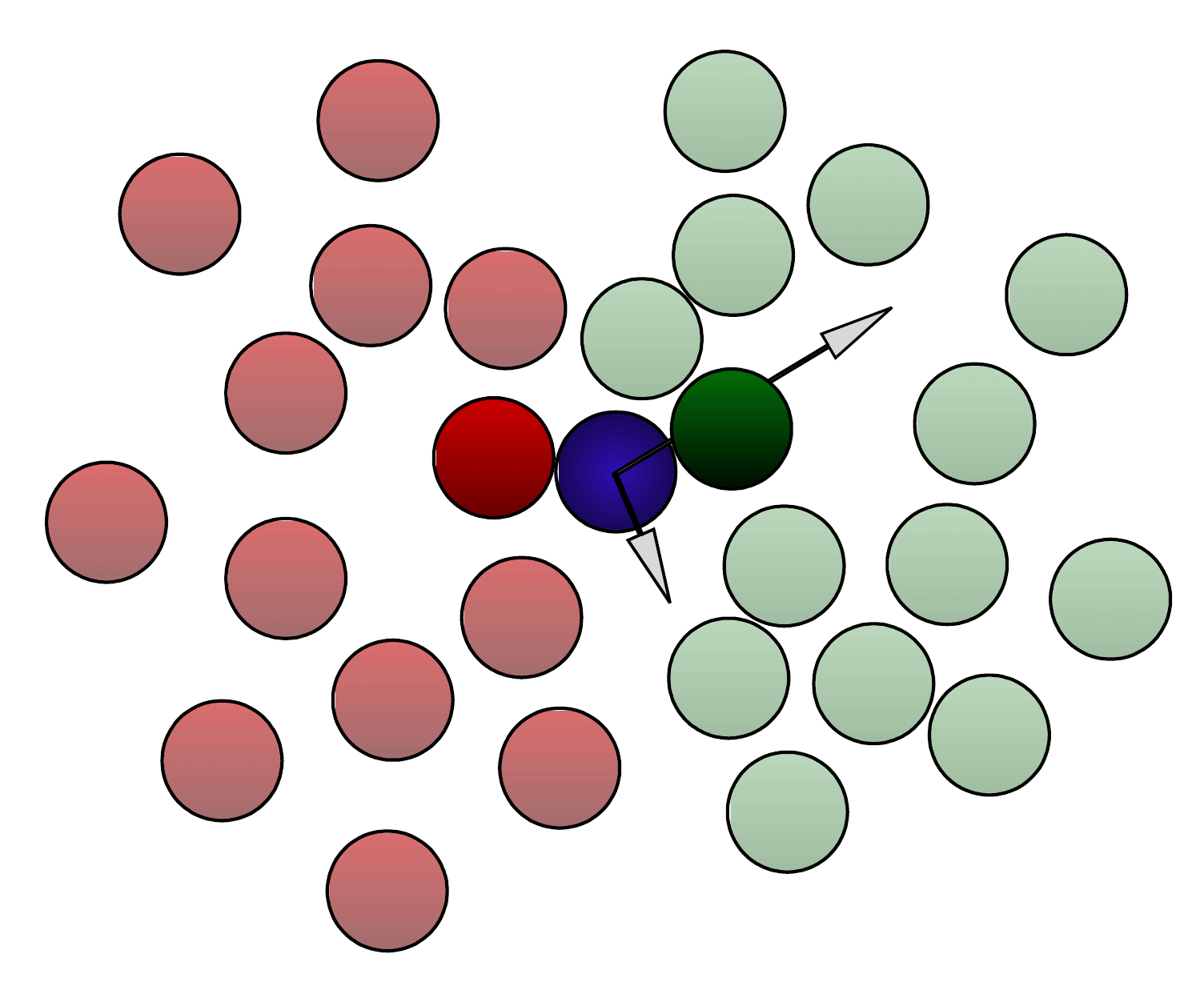}
	\put(75,55){$\mathbf e$} 
	\put(51,30){$\mathbf q$} 
	\put(30,41){$b^k$}
	\put(65,44){$f^k$}
\end{overpic} 
\end{tabular}
	\caption{\textit{Left}: Agent $k$ (dark blue)  is pushed by agent $b^k$ (dark red)  in the direction $\mathbf p=(\mathbf X_k-\mathbf X_{b^k})$. 
	\textit{Right}: Agent $k$ (dark blue) cannot be pushed, then it rearranges itself in more comfortable position following direction $\mathbf q$ which point towards a less crowded area.}
	\label{fig:pushing}
\end{figure}

Then, if the agent has no back neighbor closer than $\Dpushing$, it is not pushed and moves normally (\texttt{BLOCK 1}), see Figs.\ \ref{fig:block1a}-\ref{fig:block1b}.
Otherwise, it can have two types of dynamics:
if there is some space in front of it (i.e.\ $\df_k\geq\Dminimal$) it pushed in the direction $\mathbf p:=(\mathbf X_k-\mathbf X_{b^k})$ joining the pushing and the pushed agents (\texttt{BLOCK 2}), see Fig.\ \ref{fig:pushing}-left.
We recall here that, according to the experimental paper \cite{wang2018JSM}, we assume that the strength of the pushing force is proportional to the distance between the pushing and the pushed agents.
Let us also note that in \texttt{BLOCK 2} it is possible that an agent is pushed over another one (overlap). 
With a fine tuning, overlaps are confined to a short transient and are resolved in a few time steps. In principle, overlaps could be completely avoided by duly reducing $C$, but this is not advisable because the system greatly benefits from a certain degree of elasticity. We will come back on this point in Test 2.

If instead there is no enough space to move on, it just tries to rearrange itself by means of small steps of length $\varepsilon\Delta x$, towards a position which maximizes the minimal distance $\dn_k$, regardless of the target and of the pushing direction (\texttt{BLOCK 3}), see Fig.\ \ref{fig:pushing}-right.
\begin{rem}
	In Blocks 1 and 3, it is likely that more than one solution exists for the optimization problems. In that case the solution should be chosen at random among the possible ones (otherwise a biased drift in a specific direction would appear). 
\end{rem}

\subsection{Possible generalizations}\label{sec:generalizations}
When it comes the moment of practical applications, the assumptions that all agents are equal is an important limitation of the models. 
Likely, the model described above is easily generalizable: making the parameters $\Rmin$, $\Rmax$, $\Sref$, $\alpha$, $\varepsilon$, and $C$ be agent-dependent, one can have pedestrians variable in size, speed, response to the crowd, and behavior in crowded conditions. 
Making $\Dcomfort$ be agent-dependent, one can simulate different degree of adaptability of the pedestrians to different situations.
Moreover, it is possible to assume that pedestrians have a noncircular, more realistic shape (ellipse or similar). The price to pay is a slower and more complicated numerical code.

It is also possible to consider the presence of \textit{social groups}, i.e.\ small groups of people who want to stay close to each other at any time. 
This can be easily achieved by adding a further distance-based constraint: agents cannot move in positions too far from the members of their social group.


Finally, people can also have a different target. In this case an eikonal equation for each person must be solved. 

\section{Numerical simulations}\label{sec:numerics}
In this section we present seven numerical simulations to show the potential of the model. We have chosen three settings among the most commonly investigated ones in the literature: a corridor, a room, and a corner, with one or two targets.

The algorithm described above is not fully discrete, so a further step is needed to make it implementable. 
To simplify the solution to the optimization problems appearing in \texttt{BLOCK 1} and \texttt{3}, we consider only a finite number of values for $\theta$. 
More precisely, we evaluate only 36 possible directions of motion, one every 10$^\circ$. 
Minimization and maximization is then performed by a simple direct comparison.

The time step is chosen as $\Delta t=0.1$ and $\varepsilon=0.1$. The other default parameters are summarized in Table \ref{tab:defaultparameters}. 
\begin{table}[h!]
	\caption{Default parameters. All distances are expressed in [m], $\Sref$ in [m/s] and $C$ in [s$^{-1}]$.}
	\begin{center}
		\begin{tabular}{|c|c|c|c|c|c|c|c|}
			\hline
			$\Dcomfort$ &
			$\Dcontact$ &
			$\Dpushing$ &
			$\Dminimal$ &
			$\Sref$ &
			$\alpha$ &
			$C$ & 
			$\varepsilon$ \\ \hline
			1     & 0.5  & 0.45 & 0.4 & 0.8 & 2 & 4 & 0.1 \\ \hline
		\end{tabular}
		\label{tab:defaultparameters}
	\end{center}
\end{table}
These parameters were chosen after some discussions with practitioners, in order to be as consistent as possible, and they are tuned in order to reach reasonable maximal densities in normal and congested scenarios (Tests 1 \& 2). 
Note that, because of the assumption, made here, that pedestrians are circular shaped, parameters regarding distances necessarily represent an average of real measures (for example, the actual contact distance clearly depends on the orientation of the torso).

Table \ref{tab:allparameters}, instead, summarize the parameters which change in Tests 1--6 with respect to the default ones.
\begin{table}[h!]
	\caption{Choice of parameters in Tests 1--6. All distances are expressed in [m], $\Sref$ in [m/s] and $C$ in [s$^{-1}]$. Missing values must be considered as unchanged with respect to default ones.}
		\begin{center}
	\begin{adjustbox}{width=1\textwidth}
	\begin{tabular}{|l|c|c|c|c|c|c|c|c|c|}
		\hline
		Test & $N$ &
		$\Dcomfort$ &
		$\Dcontact$ &
		$\Dpushing$ &
		$\Dminimal$ &
		$\Sref$ &
		$\alpha$ &
		$C$ & 
		$\varepsilon$ 
		\\ \noalign{\hrule height2pt}
  		1a   & 400 &   &  & & & & &  & \\ \hline
		1b &  	400	  & 1.2  & 0.6  & 0.54 & 0.48 &  &  & & \\ \hline
		1c &  400		  & 0.8  & 0.4  & 0.36  & 0.32  &  &  &  & \\  \hline
		1d  &  	400	   &              &             &              &              &  & 0.5 & & \\ 
		\noalign{\hrule height2pt}
		2a   &400 & & & & & & &  & 0\\ \hline
		2b &400 &  &  &  &  &  &  & &    \\ \hline
		2c & 400 &  &  &  &  &  &  & & 0.5\\ \hline
		2d & 400 &  &  &  &  &  &  & & 2\\ \hline
		2e & 400 &  &  &  &  &  &  & 1 & 0\\ \hline
		2f & 400 & &  &  &  &  & & 0.1 & \\ 
		\noalign{\hrule height2pt}
		3   & var &  & 0.4 & 0.35 & 0.3 &  &  &   &  \\ 
		\noalign{\hrule height2pt}
		4   & 2000 & 0.6 & 0.6 &  & & & & & \\ 
		\noalign{\hrule height2pt}
		5   & 100 &  &  &  &  & 0.6 &  &  &  \\ 
		\noalign{\hrule height2pt}
		6   & 20 & 0.6 &  &  &  & 1 & & & \\ 
		\hline
	\end{tabular}
\end{adjustbox}
	\end{center}
\label{tab:allparameters}
\end{table}

The fastest path to the target is computed analytically by hand, except for Test 5 where the eikonal equations (\ref{ee}) and (\ref{ee-modif}) are used.
Eikonal equation is solved by means of a semi-Lagrangian scheme \cite{falconebook}, with a space step equal to 0.1. 

In the figures, pedestrians are drawn as blue circles of radius $\Dminimal$.
If they are pushed ($\db_k<\Dpushing$) they are drawn in red. 


\subsection{Test 1: Closed corridor}
In this first test we simulate 400 people walking through a corridor of size 60 m $\times$ 10 m, and approaching a closed gate (target) located at the rightmost side and as wide as the corridor. People in front stop at the gate and then a queue progressively extends backwards.
One could expect that once the people have stopped (just in front of the gate or behind other stopped pedestrians) they stay still forever. Actually this is not what happens: people sometimes start moving again, as they are `pressed' by the people behind them. 
This is a very nice self-regulating effect which is quite hidden in the algorithm and comes from the fact that people in \texttt{BLOCK 1} move taking into account people in front only. 
As a consequence, it can happen that an agent gets close to another agent who is behind it, causing the decrease of its accepted distance $\da$ and then a restarting to get closer to the crowd in front. 

\subsubsection*{Test 1a: Default parameters}
\begin{figure}[h!]
	\centering
	 	\begin{overpic}[width=0.9\textwidth]{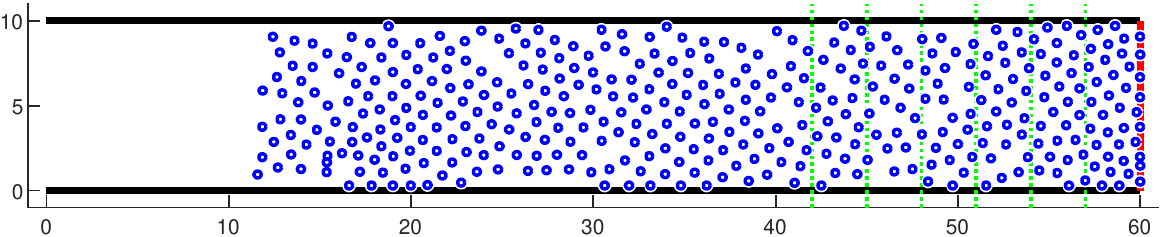}
	 		\put(71,19){1} \put(76,19){2} \put(81,19){3} 
	 		\put(86,19){4} \put(90.5,19){5} \put(95,19){6} 
	 	\end{overpic}\\
	 \includegraphics[width=0.9\textwidth]{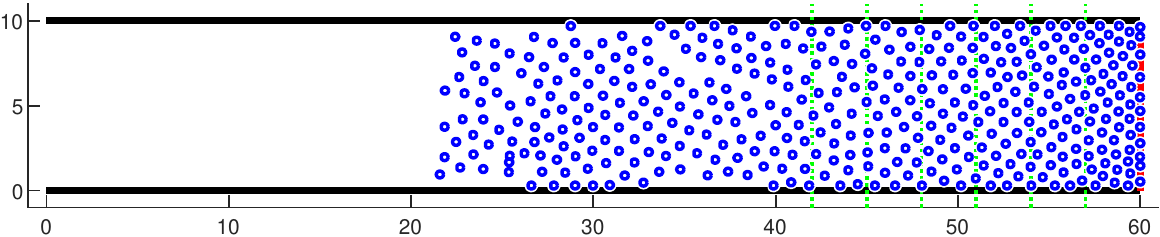}\\
	 \includegraphics[width=0.9\textwidth]{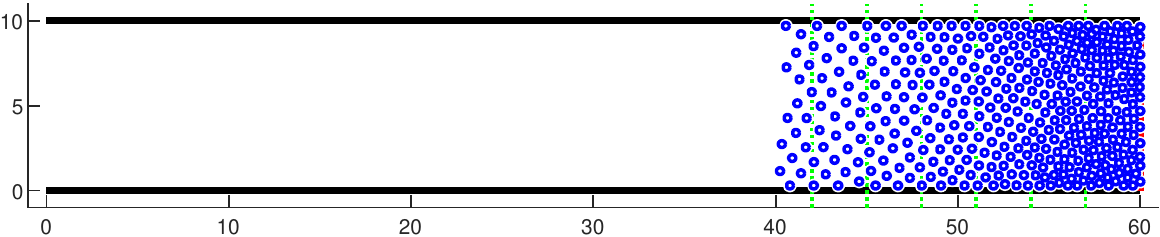}
	\caption{Test 1a. \textit{From top to bottom}: three snapshots of the simulation at $n=75, 200, 700$. Six regions (green strips) near the gate are marked and numbered, they will be used later on in Fig.\ \ref{fig:T1_density}.
	Animated simulation available  \mbox{\href{www.emilianocristiani.it/attach/pedestrians2022-T1.mp4}
	{here}}.	
	Numerical code available  \mbox{\href{http://www.emilianocristiani.it/codes/TEST_PAPER_1a_public.zip}{here}}.
	\label{fig:T1a_corridor}
	}
\end{figure}
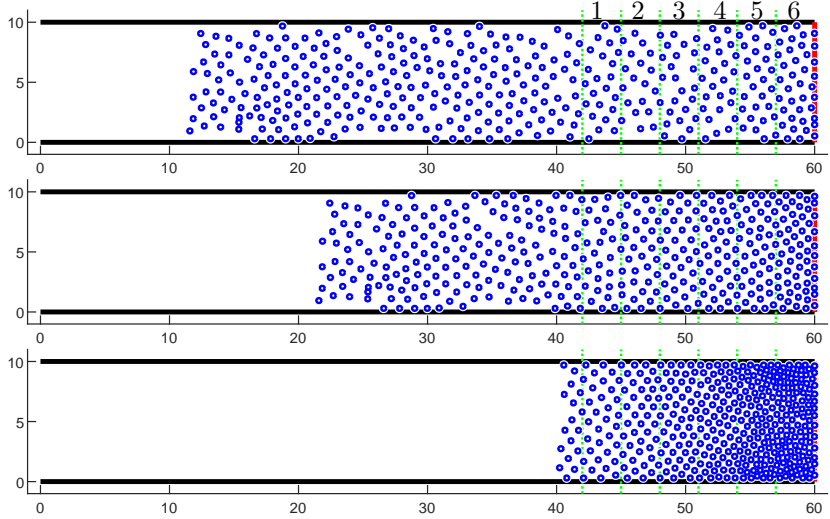

Fig.\ \ref{fig:T1a_corridor} shows three snapshots of a simulation obtained with default parameters, see Table \ref{tab:defaultparameters}.
Initial positions are random. In the last snapshot the crowd has reached an equilibrium and it is still. Remarkably, the crowd in equilibrium has a \textit{nonhomogeneous density}, being higher near to the target. This effect is caused by the restarting mechanism mentioned before.
Let us note here that the original optimal step model is not able to reproduce this phenomenon, and an \textit{ad hoc} modification is needed to get it, see \cite{vonsivers2015TRB}.

\begin{figure}[h!]
	\centering
	\begin{tabular}{cc}
	\includegraphics[width=0.49\textwidth]{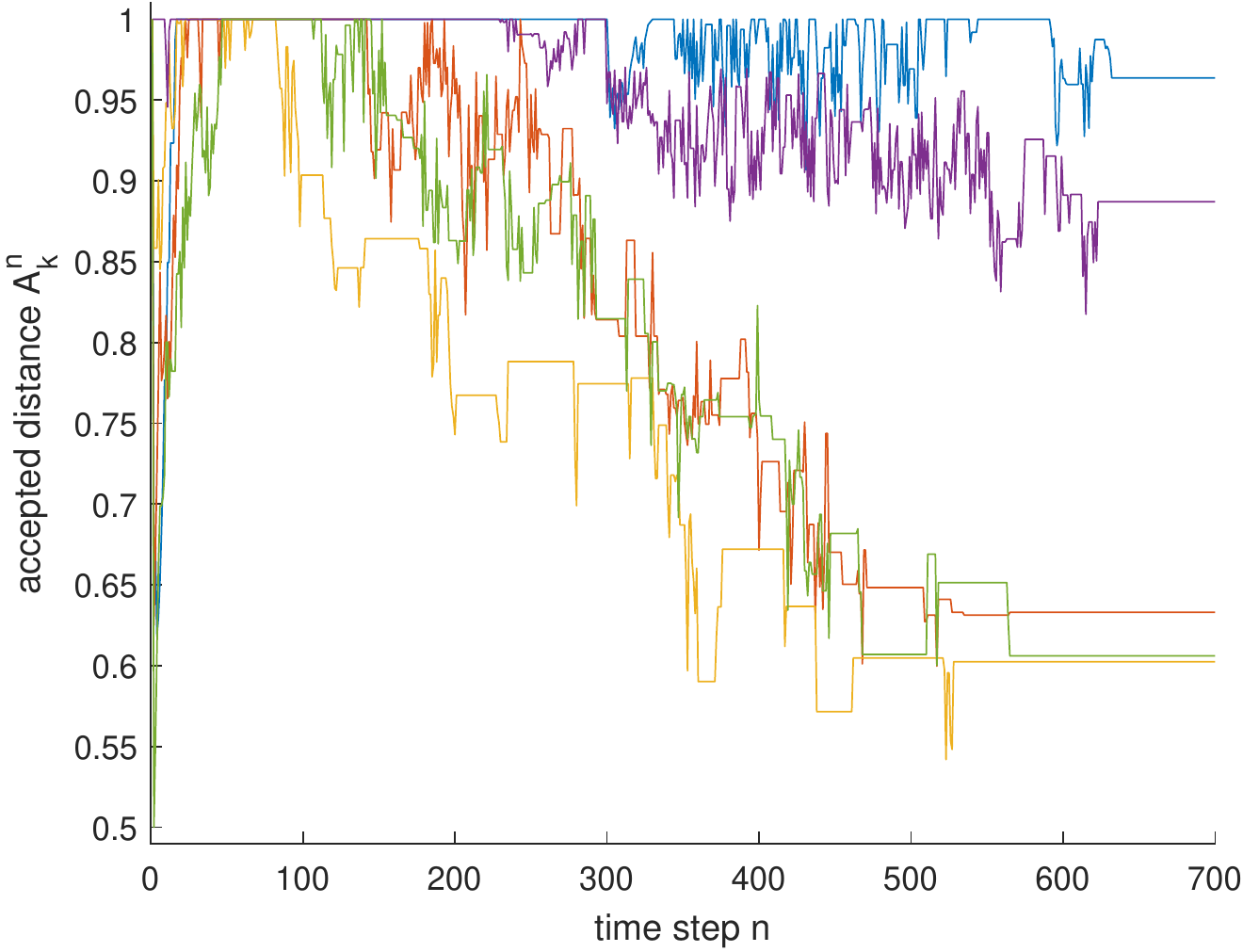} &
	\includegraphics[width=0.49\textwidth]{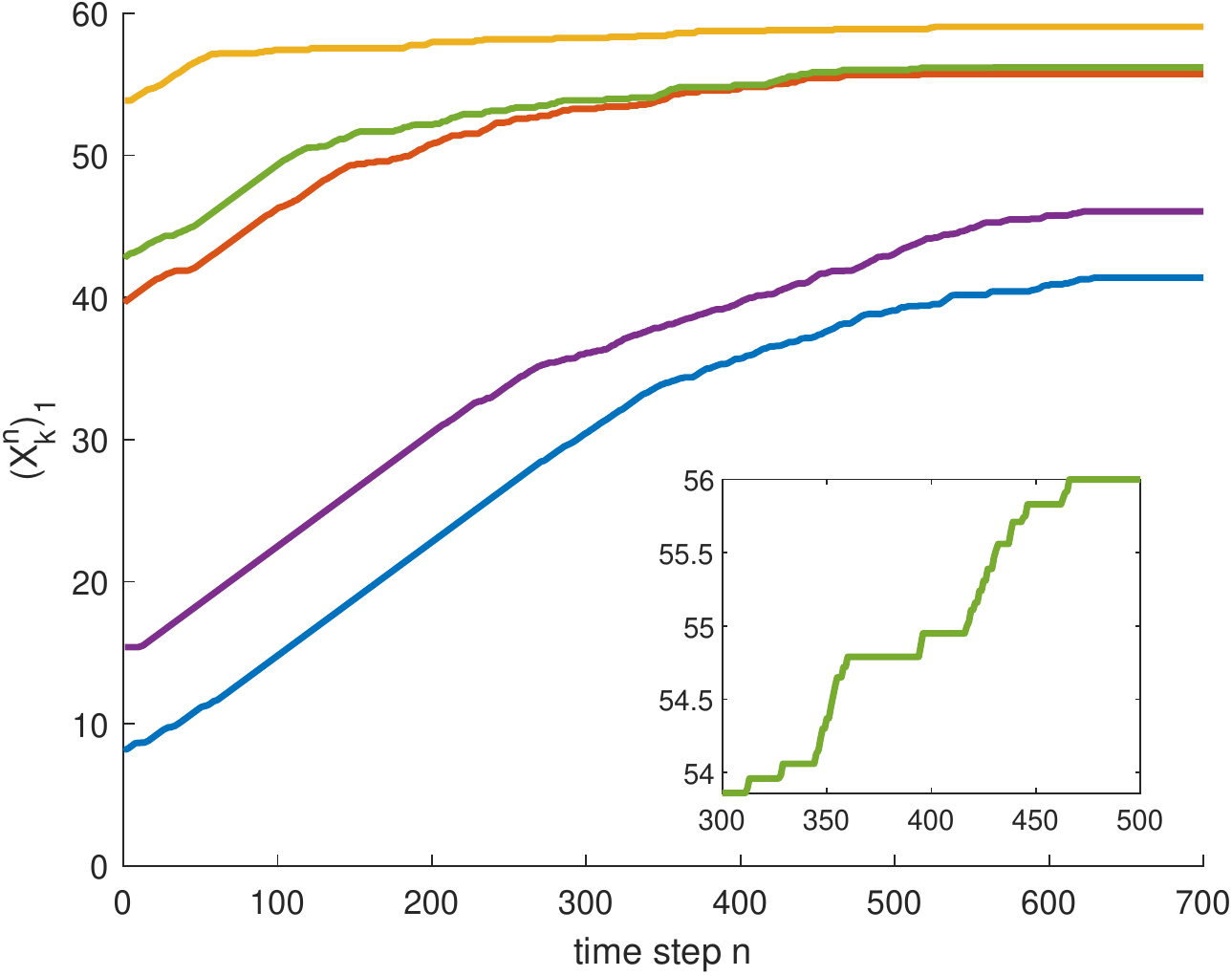} 
\end{tabular}
	\caption{Test 1a. \emph{Left}: the function $n\to\da^n_k$ for 5 random agents. We see that the functions both increase and decrease, and stabilizes at the end once the equilibrium is reached.
	\emph{Right}: the function $n\to(\mathbf X^n_k)_1$ for the same 5 agents. The slope of the graphs corresponds to the velocity of the agents. We see that they slow down approaching the gate, by means of a stop \& start mechanism (see inset).
Comparing the two figures we also note that the agents close to the gate (yellow, green, red) have a lower value of $\da$, as expected from Fig.\ \ref{fig:T1a_corridor}.}
	\label{fig:T1_Ak_and_Xk}
\end{figure}

Fig.\ \ref{fig:T1_Ak_and_Xk} shows two properties of five randomly chosen agents. In particular, we show the time evolution of their accepted distance $\da^n_k$ and the time evolution of their first (horizontal) coordinate of the position $\mathbf X^n_k$, related in the obvious manner to the distance to target.

\begin{figure}[h!]
	\centering
	\includegraphics[width=0.9\textwidth]{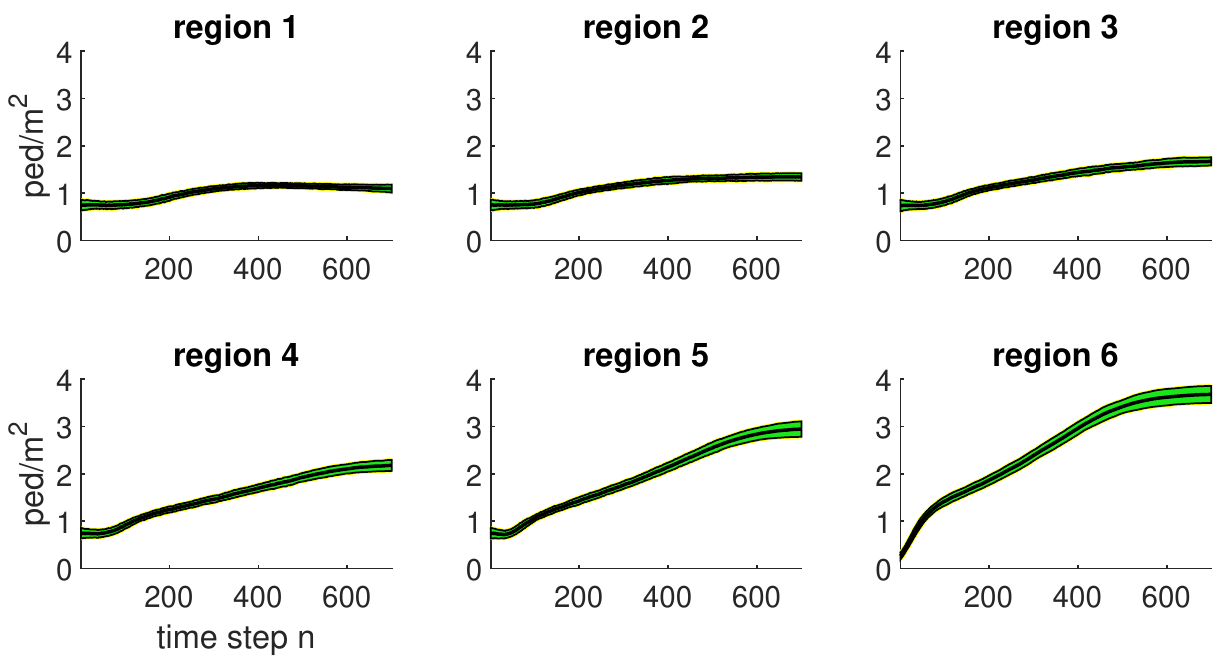}
	\caption{Test 1a. Density as a function of time step $n$ in the 6 regions marked in Fig.\ \ref{fig:T1a_corridor} (the green strip represents the area $\mu\pm\sigma$ over 300 runs). 
	We see that at final time the density is not constant and reaches the maximum (3.7 ped/m$^2$) in the region closest to the target.}
	\label{fig:T1_density}
\end{figure}

Fig.\ \ref{fig:T1_density} shows the time evolution of the density of people in the 6 regions marked in Fig.\ \ref{fig:T1a_corridor} by the green vertical dotted lines. Results are averaged over 300 runs.

\subsubsection*{Tests 1b--1d: Sensitivity analysis}
Here we vary some parameters of the model in order to show their role in the dynamics.

\begin{figure}[h!]
	\centering
		\includegraphics[width=0.6\textwidth]{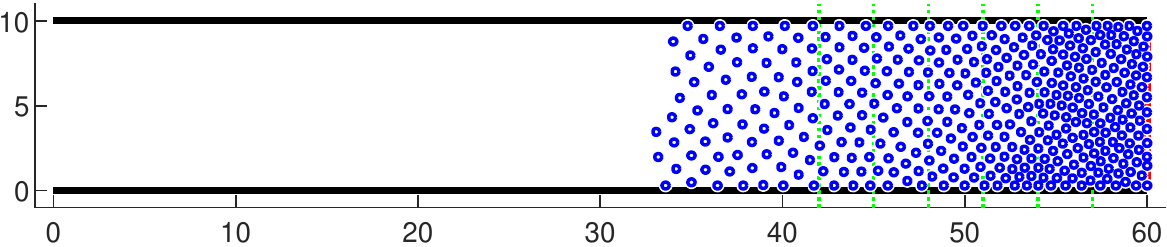} \\
		\includegraphics[width=0.6\textwidth]{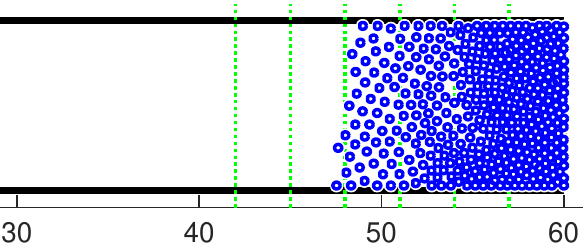} 	
	\caption{Tests 1b \& 1c.  Equilibrium configuration with all distances $D$'s increased (\emph{top}) and decreased (\emph{bottom}) by 20\%. Densities in the rightmost regions are 2.6 and 5.1 ped/m$^2$, respectively.}
	\label{fig:T1sensitivity_D}
\end{figure}

Fig.\ \ref{fig:T1sensitivity_D} shows the final configurations reached by the crowd obtained by varying all the distances $D$'s by $\pm$20\%. 
The results must be compared with Fig.\ \ref{fig:T1a_corridor}-bottom. 
As expected, the crowd expands/shrinks accordingly, still preserving the nonhomogeneity in the density. 

\begin{figure}[h!]
	\centering
	\includegraphics[width=0.6\textwidth]{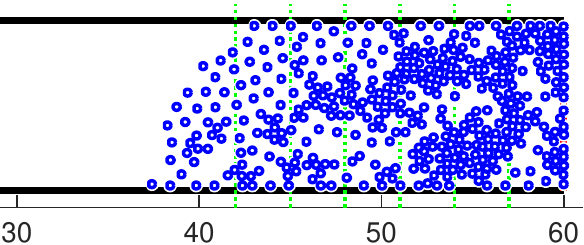} 	
	\caption{Test 1d. Equilibrium configuration with $\alpha$ decreased from 2 to 0.5.}
	\label{fig:T1sensitivity_alpha}
\end{figure}

Fig.\ \ref{fig:T1sensitivity_alpha} shows instead the final configuration reached by decreasing $\alpha$ from 2 to 0.5. 
One can see that the solution degrades since spurious blockages appear. 
Instead, increasing $\alpha$ has a minor effect in this case, this can be explained considering the fact that $\da_k^{(n+1)}$ is in any case bounded by above by $\Dcomfort$, see Fig.\ \ref{fig:diagrammadiflusso}.

\clearpage
\subsection{Test 2: Closed corridor with pushing}
\begin{figure}[h!]
	\centering
	\begin{overpic}[width=0.46\textwidth]{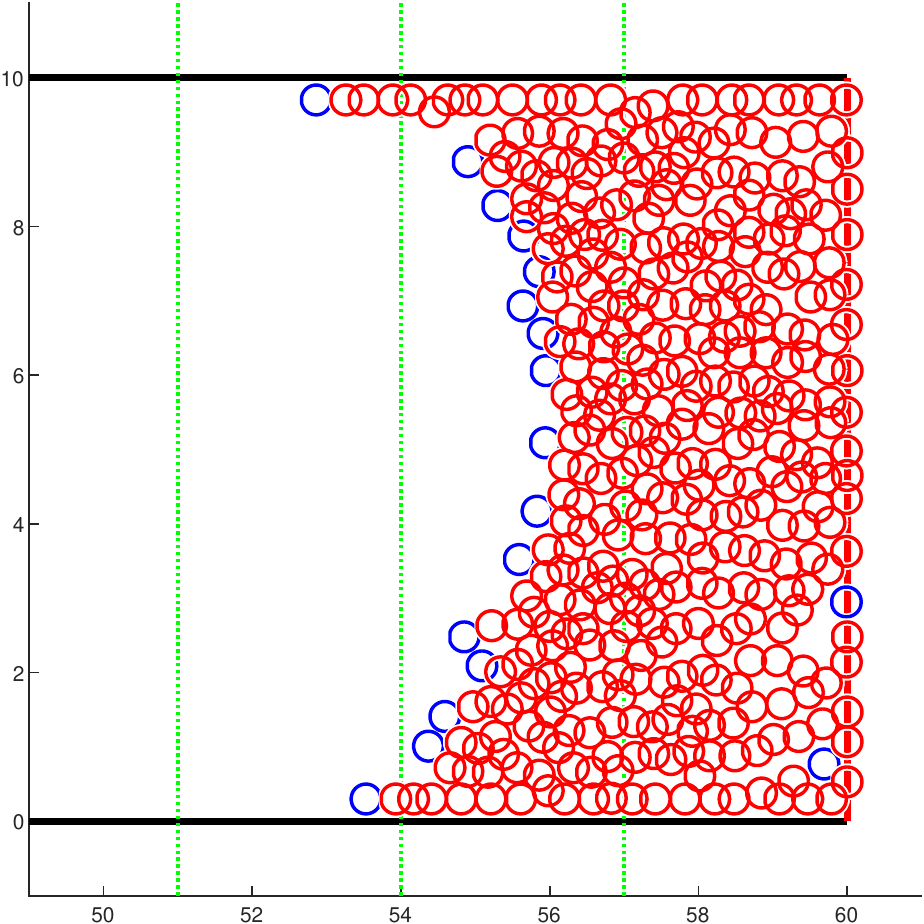}
		\put(10,70){$C=4$} \put(10,60){$\varepsilon=0$} 
	\end{overpic}\qquad
	\begin{overpic}[width=0.46\textwidth]{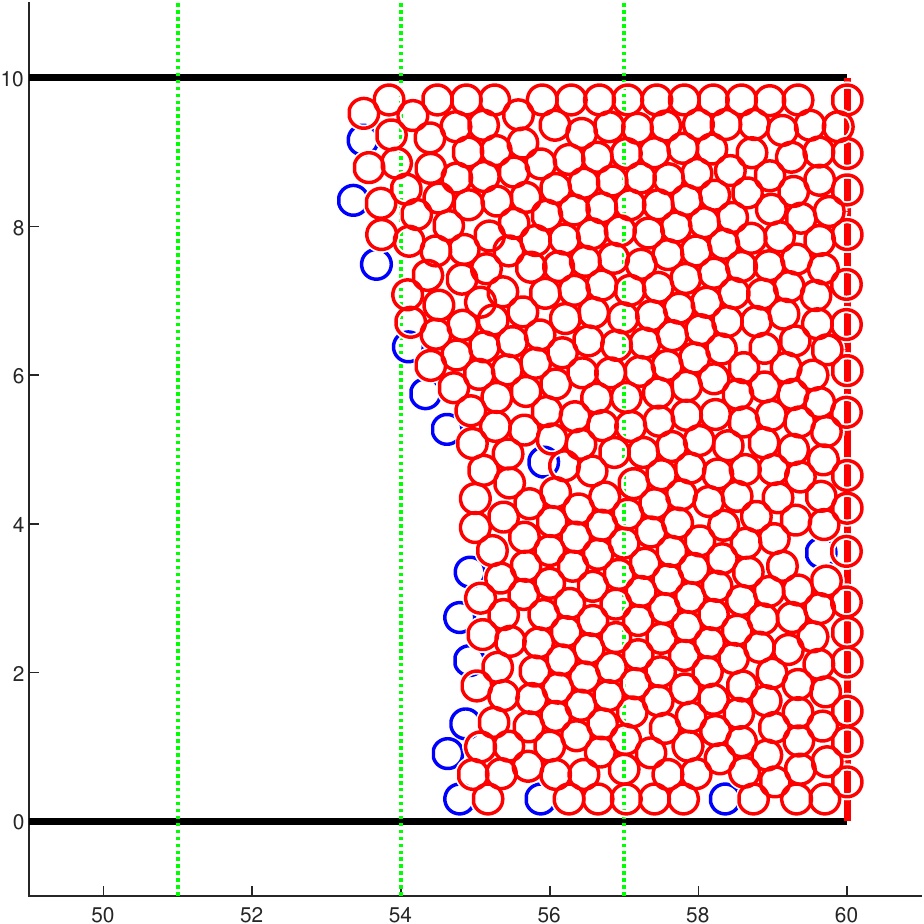}
		\put(10,70){$C=4$} \put(10,60){$\varepsilon=0.1$} 
	\end{overpic}\\
	\begin{overpic}[width=0.46\textwidth]{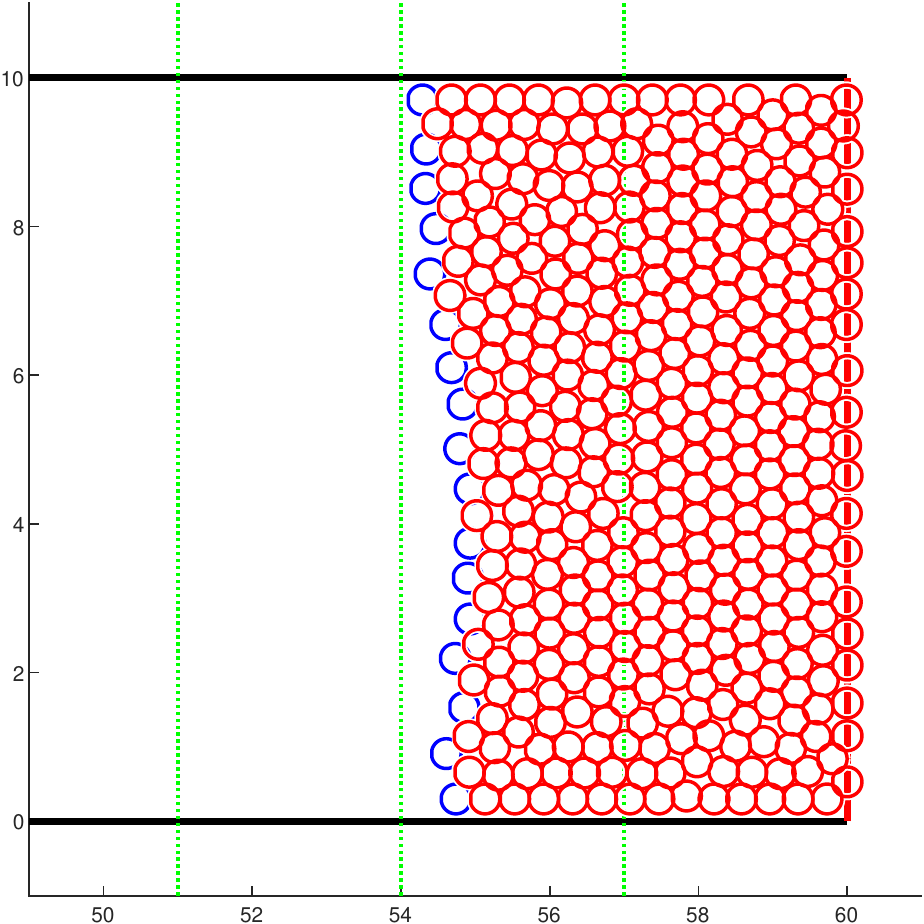}
		\put(10,70){$C=4$} \put(10,60){$\varepsilon=0.5$} 
	\end{overpic}\qquad
	\begin{overpic}[width=0.46\textwidth]{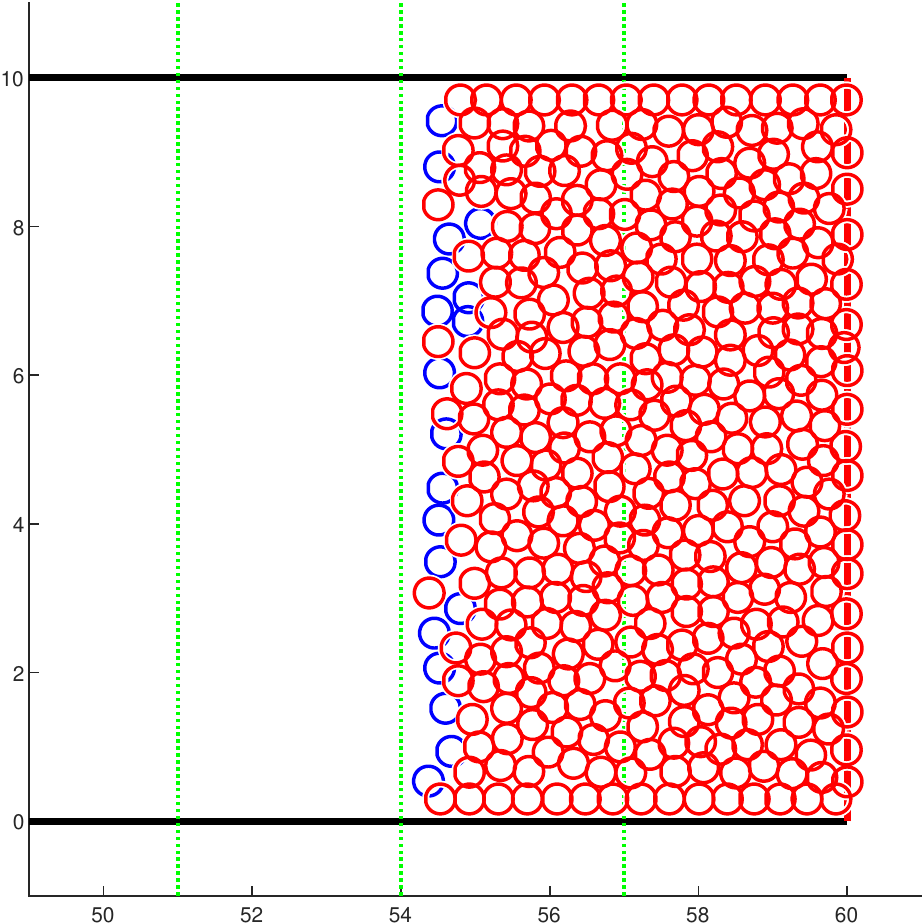}
		\put(10,70){$C=4$} \put(10,60){$\varepsilon=2$} 
	\end{overpic}\\
	\begin{overpic}[width=0.46\textwidth]{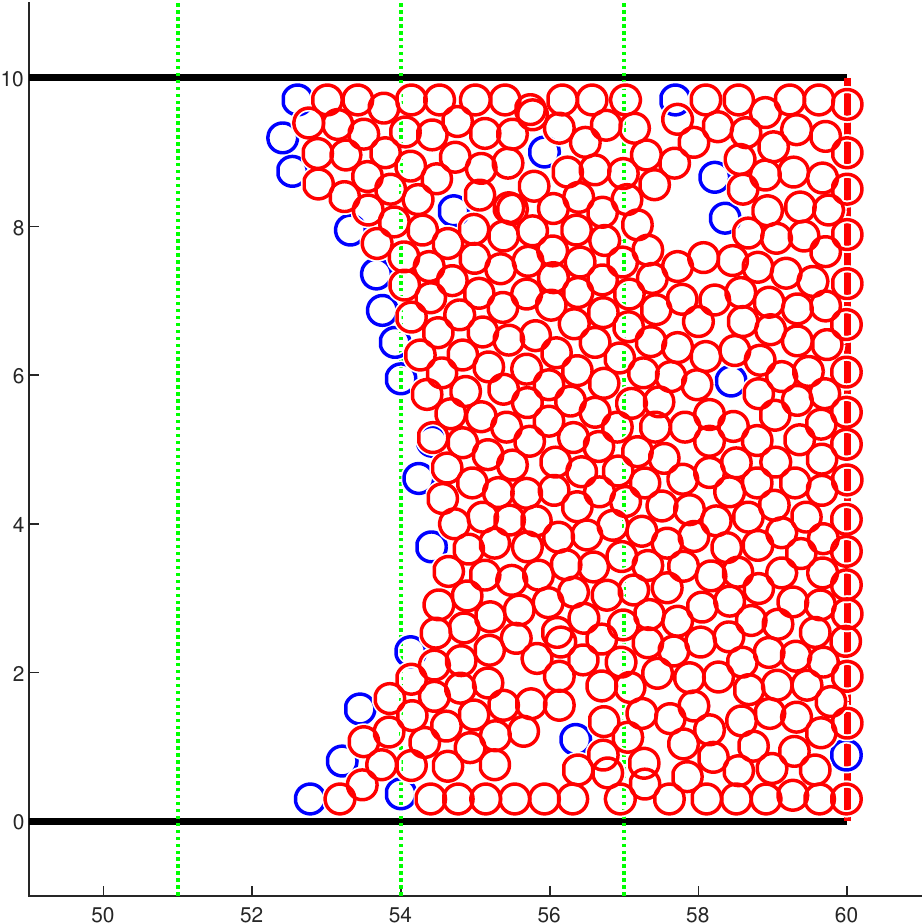}
		\put(10,70){$C=1$} \put(10,60){$\varepsilon=0$} 
	\end{overpic}\qquad
	\begin{overpic}[width=0.46\textwidth]{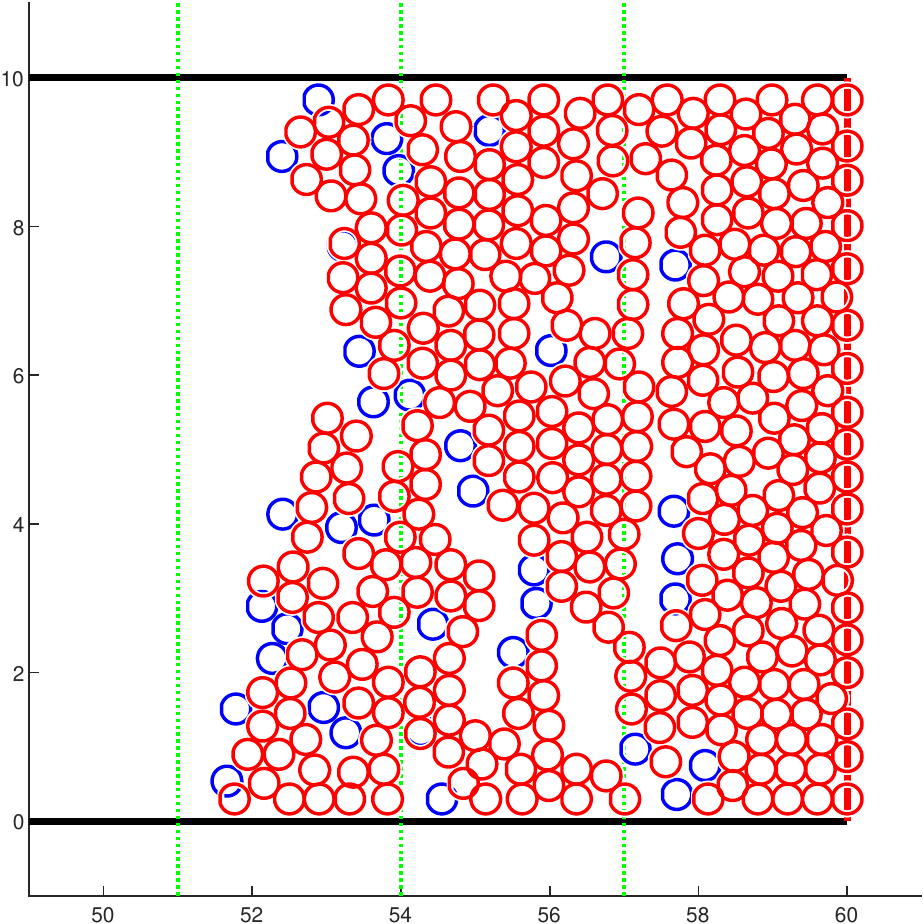}
		\put(10,70){$C=0.1$} \put(10,60){$\varepsilon=0.1$} 
	\end{overpic}
	\caption{Tests 2a--2f. Crowd at equilibrium with $\da_k=\Dminimal$ $\forall k$. Red agents are the pushed ones. 
	\emph{2a, Top-left}: Pedestrians overlap because of the pushing force, and overlapping is never resolved because of the removal of \texttt{BLOCK 3}.
	\emph{2b, Top-right}: default parameters. Pedestrians occasionally overlap but overlapping is resolved in short time. The maximal density reached in the rightmost region at equilibrium is 6.8 ped/m$^2$. 
	\emph{2c, Mid-left}: Pedestrians occasionally overlap during the simulation (not shown) but at the end they reach a very ordered pattern with maximal density of 7.3 ped/m$^2$.
	\emph{2d, Mid-right}: Pedestrians overlap during the simulation, and the overlapping slightly persists at equilibrium.
	\emph{2e, Bottom-left}: Pedestrians do not overlap since they are less pushed, but they are not able to fill the free space around them because of the removal of \texttt{BLOCK 3}, leaving spurious holes.  
	\emph{2f, Bottom-right}: Even if \texttt{BLOCK 3} is active and well tuned, pedestrians are not able to fill the whole space since the pushing force is too small.
}
\label{fig:T1_corridor_all_panicked}
\end{figure}
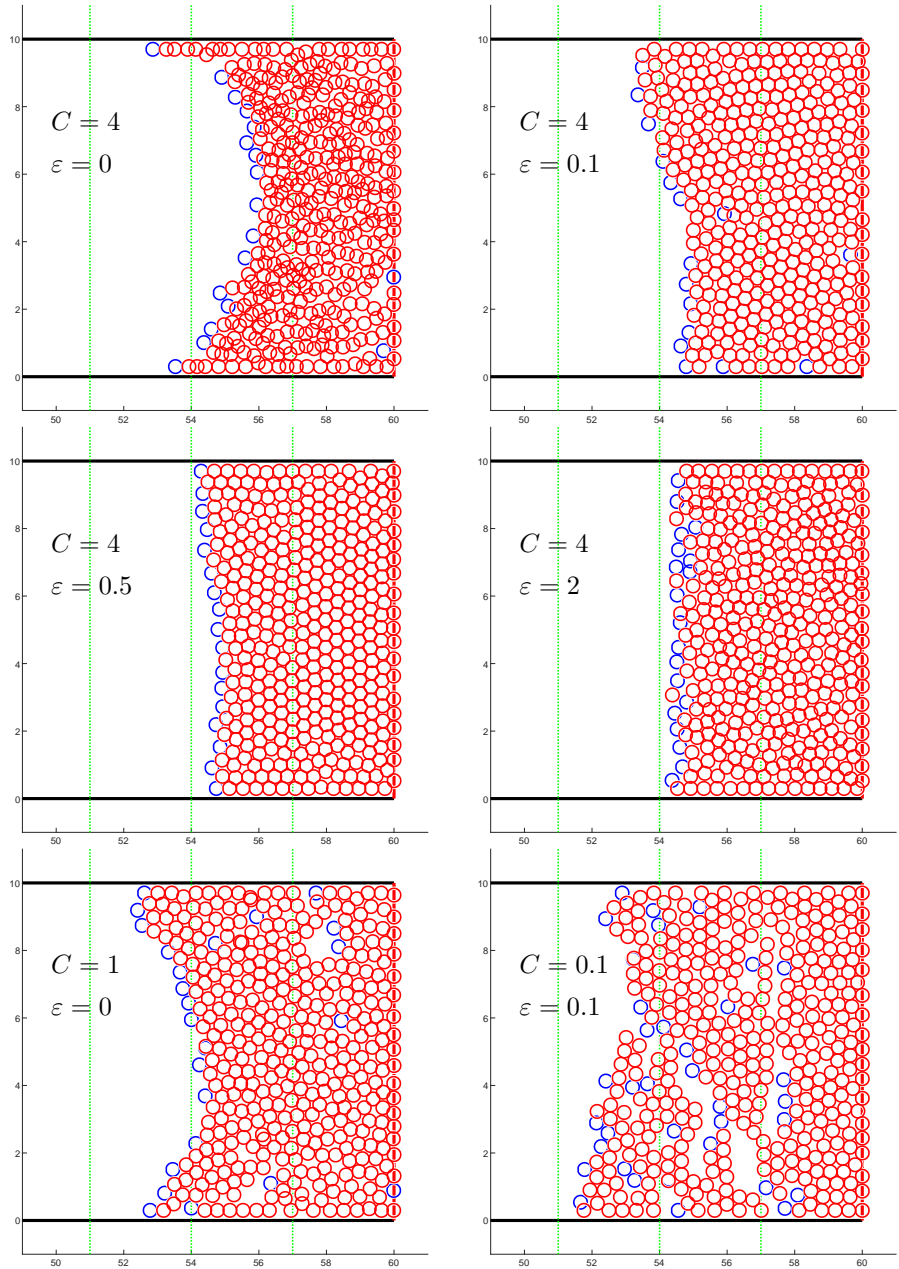

To show the effect of pushing behaviour (\texttt{BLOCKS 2-3}) we consider again the scenario of Test 1a. We extend the final time to $n=1000$ and, at $n=700$, we reduce the accepted distance of \textit{all} agents to the minimum, i.e.\ $\da_k=\Dminimal$ $\forall k$. 
All agents start pushing and being pushed, until they rearrange in the stationary configuration with maximal density. 
Fig.\ \ref{fig:T1_corridor_all_panicked} collects the results for six choices of the parameters related to the pushing behaviour, namely $C$ and $\varepsilon$. 
We have also included the case $\varepsilon=0$, which corresponds to a complete removal of \texttt{BLOCK 3}. This is done to better highlight the role of this block.

From the figure we deduce that the choice of the default parameters is the best compromise if one wants to keep the overlap to a minimum and, at the same time, reaching high densities by filling all the free space between people.

\clearpage
\subsection{Test 3: Circular corridor and fundamental diagram}\label{sec:T2}
In this test we aim at computing the fundamental diagram as it naturally results from the model. We recall that the model does not require a fundamental diagram as a parameter, but macroscopic quantities as average velocity, density and flux can estimated \textit{a posteriori} from the simulation.
\begin{figure}[b!]
	\centering
	\includegraphics[width=0.49\textwidth]{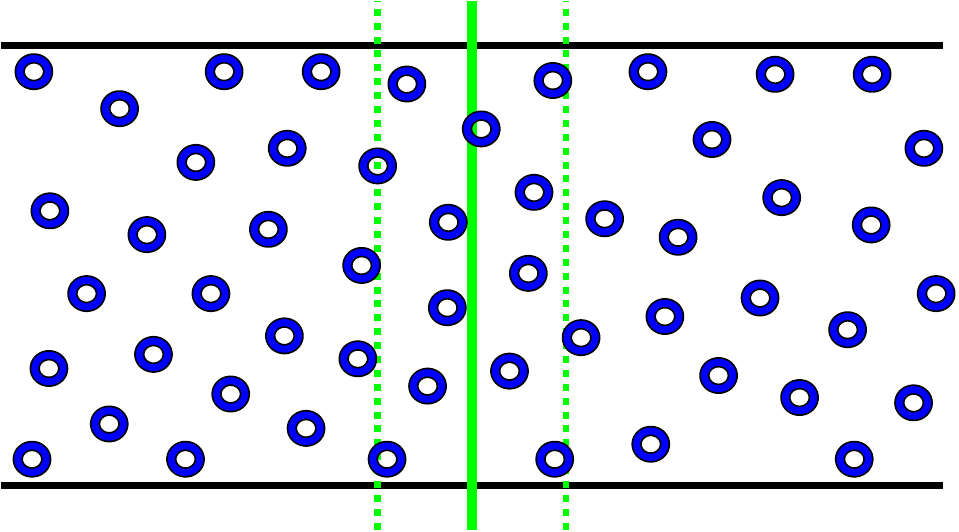} 
	\includegraphics[width=0.49\textwidth]{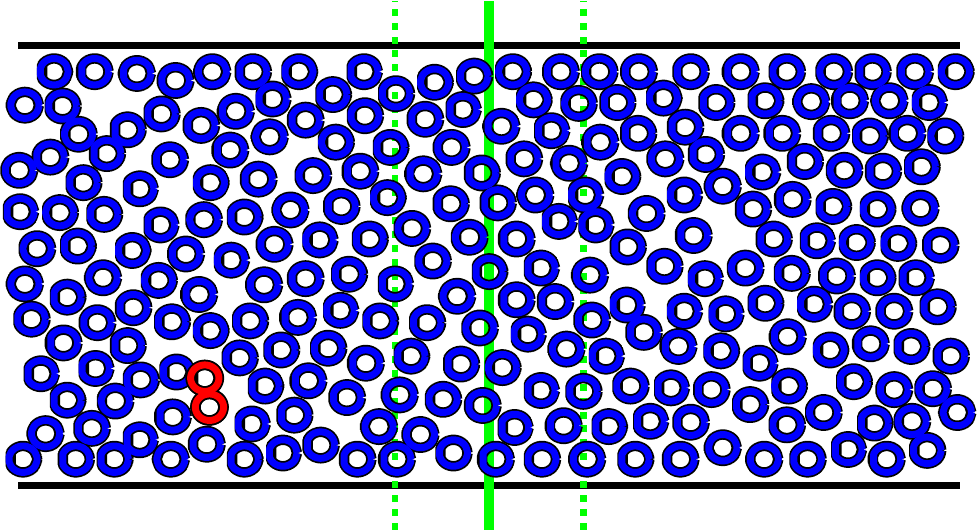}\\ 
	\includegraphics[width=0.49\textwidth]{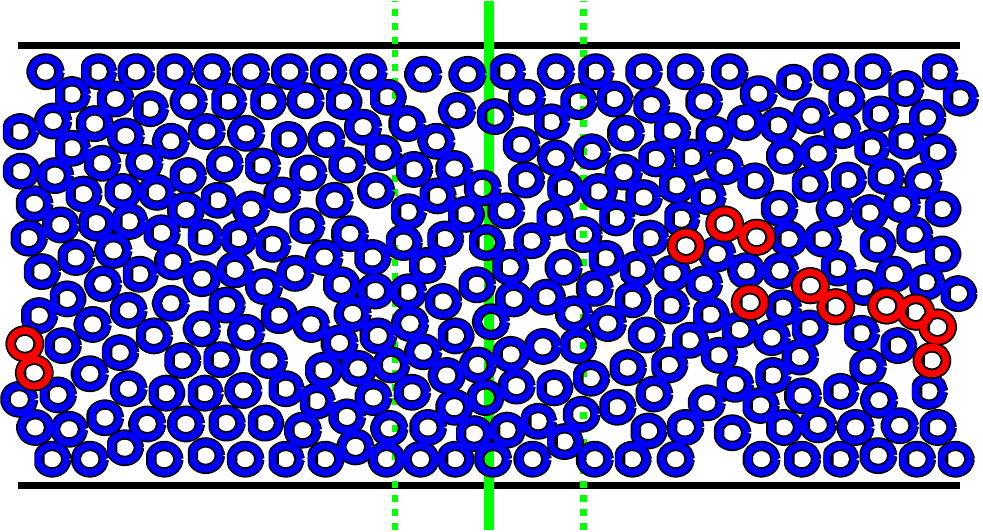} 
	\includegraphics[width=0.49\textwidth]{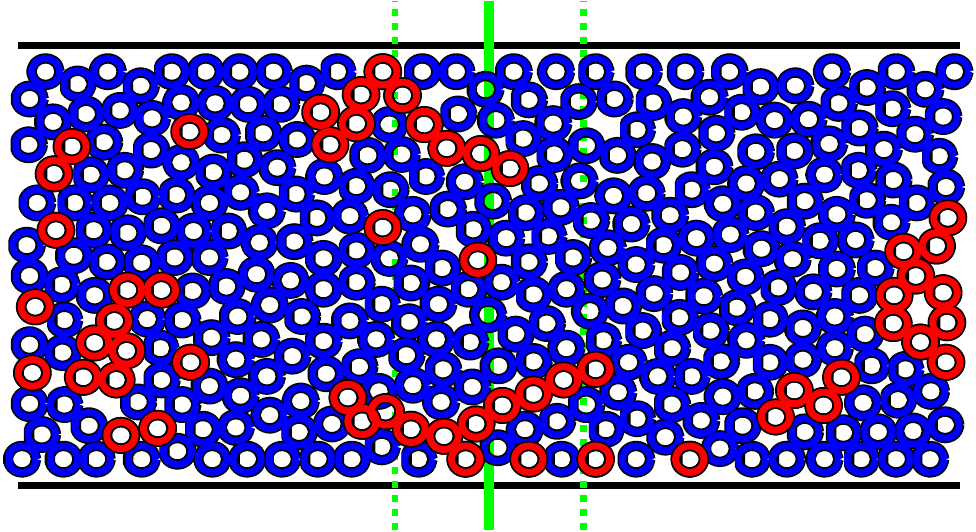}\\ 
	\includegraphics[width=0.49\textwidth]{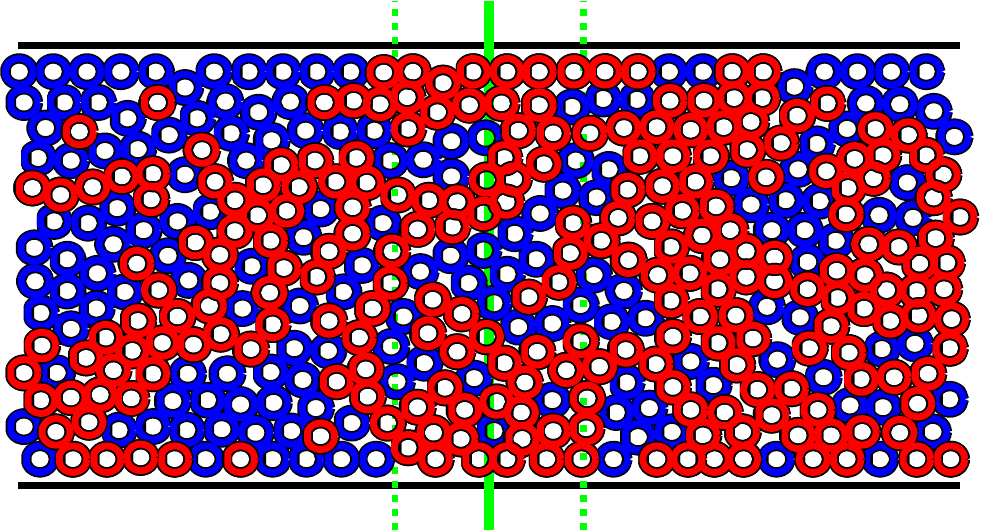} 
	\includegraphics[width=0.49\textwidth]{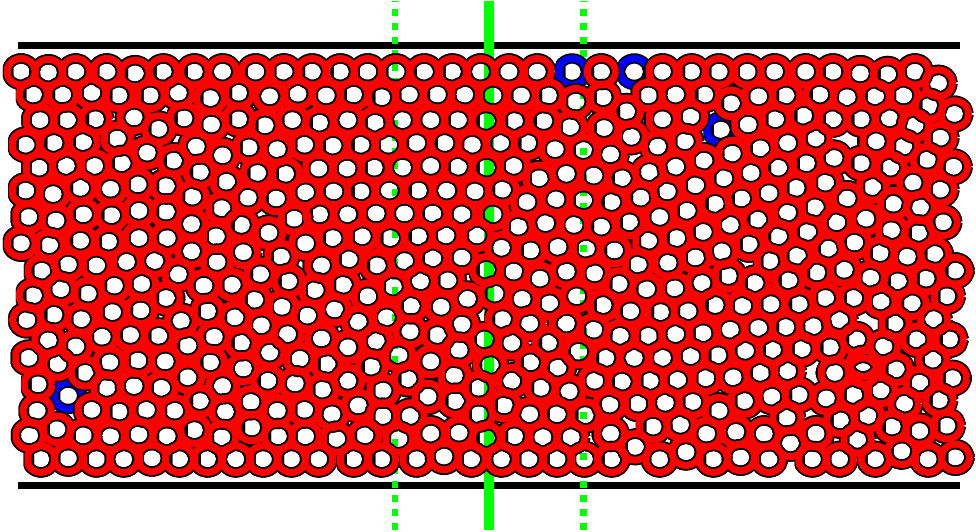} 
	\caption{Test 3. \textit{From left to right, top to bottom}: snapshots of the simulation as the density increases. Red circles represent pushed agents. 
		We observe six conditions: 
		1. Pedestrians are in a free flow regime and move at maximal speed ($n=500$). 
		2. They start to slow down and the maximum flux is reached ($n=2500$). 
		3. They stop almost completely and start pushing ($n=3100$).
		4. They restart moving, slowly ($n=3500$).
		5. They reach a second peak of flux ($n=4000$).
		6. They stop again ($n=5500$).
	Animated simulation available  \href{www.emilianocristiani.it/attach/pedestrians2022-T2.mp4}{here}. 
	Numerical code available  \mbox{\href{http://www.emilianocristiani.it/codes/TEST_PAPER_3_public.zip}{here}}.	
	}
	\label{fig:T2_sim}
\end{figure}

In this test we have slightly decreased some distances to better fit the experimental fundamental diagram reported in \cite{helbing2007PRE}, which consider the case of a highly congested scenario.
We have 
$\Dcontact=0.4$, 
$\Dpushing=0.35$, 
$\Dminimal=0.3$, and 
$\Delta t=0.05$.

Pedestrians move rightward in a 10 m $\times$ 5 m corridor with periodic boundary conditions (i.e.\ when they reach the rightmost boundary they are teletransported at the leftmost boundary). 
Every 10 time steps a new pedestrian appears in the corridor in a random position. 
The choice of a rectangular corridor (instead of circular) and the random ingress of people directly injected in the corridor (instead of from a lateral door) is due to the need of assuring the conditions as symmetric and homogeneous as possible, also with respect to lateral displacements. 
Obviously it is possible that, when a new agent appears, it overlaps with another already in the corridor. This is not a real issue since the overlapping conditions is resolved automatically in a few time steps, since the agent behind stops.

Fig.\ \ref{fig:T2_sim} shows some snapshots of the simulation as density increases.

\begin{figure}[t!]
	\centering
	\includegraphics[width=0.49\textwidth]{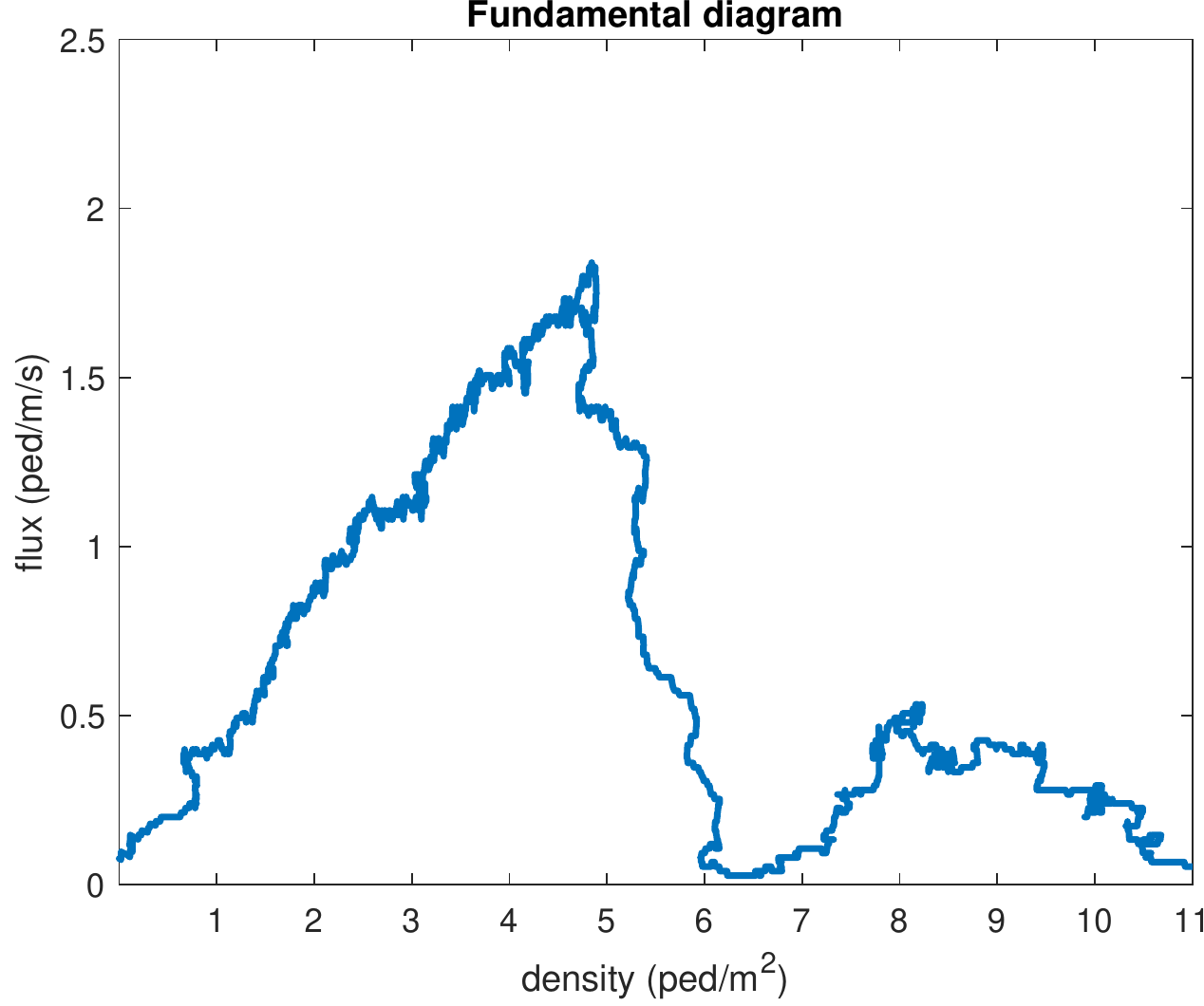}
	\includegraphics[width=0.50\textwidth]{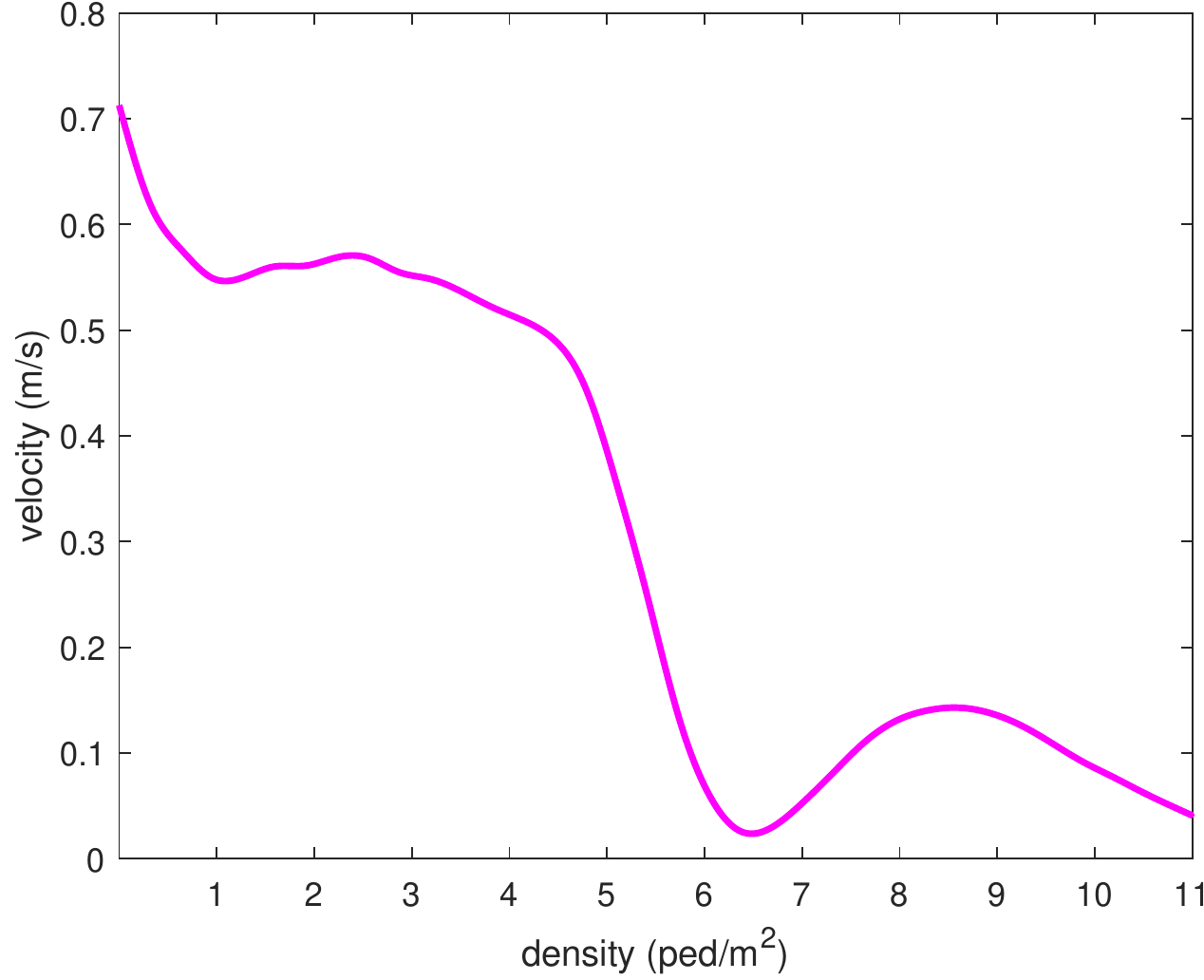}
	\caption{Test 3. \textit{Left}: flux as a function of density. The double hump is visible and it can be compared with that reported in \cite{helbing2007PRE}. 
			The second hump is clearly due to the pushing behavior, and it vanishes around 11 ped/m$^2$. \textit{Right}: velocity as a function of density. Double hump is visible here too.}
	\label{fig:T2_FD}
\end{figure}

Fig.\ \ref{fig:T2_FD} shows the relationships between flux and density (fundamental diagram) and between velocity and density.  
Regarding fundamental diagram, at any time step $n$ of the previous simulation, the density is computed by counting how many pedestrians are in the strip limited by $x=4$ and $x=6$ (dotted green lines in Fig.\ \ref{fig:T2_sim}), while flux is computed by counting how many people cross the line $x=5$ (solid green line in Fig.\ \ref{fig:T2_sim}) in that time step ($n\to n+1$). Both quantities are then smoothed by a moving average for better understanding and visualization.
Regarding velocity, instead, the procedure is a bit more complex: 
At any time step $n$ of the previous simulation, we found the set $\hat K^{n}$ of agents who are at distance more than 1 m from the four sides of the corridor. Then, at any time step $n$ and for all those agents $k\in \hat K^n$ we compute: 
\begin{enumerate}
	\item The horizontal velocity $V_k^{n}:=\frac{(\mathbf X_k^{n+1}-\mathbf X_k^{n})_1}{\Delta t}$, which is then smoothed by a moving average in time for any fixed $k$.
	\item The density $\rho_k^{n}$, counting how many people are within distance 1 m from agent $k$. 
\end{enumerate}
Finally we have fitted the point cloud $\{\rho_k^{n},V_k^{n} \}_{k\in\hat K}$ by means of the Matlab's built-in function `smoothing spline'.

The double hump (second peak) in the fundamental diagram is perfectly visible and it is directly comparable with that reported in \cite{helbing2007PRE}.
We think it is remarkable how such a minimal model can reproduce a complex dynamics like this.
%
%
%
%
%
%

\subsection{Test 4: Material wave}
\begin{figure}[h!]
\centering
\includegraphics[width=0.49\textwidth]{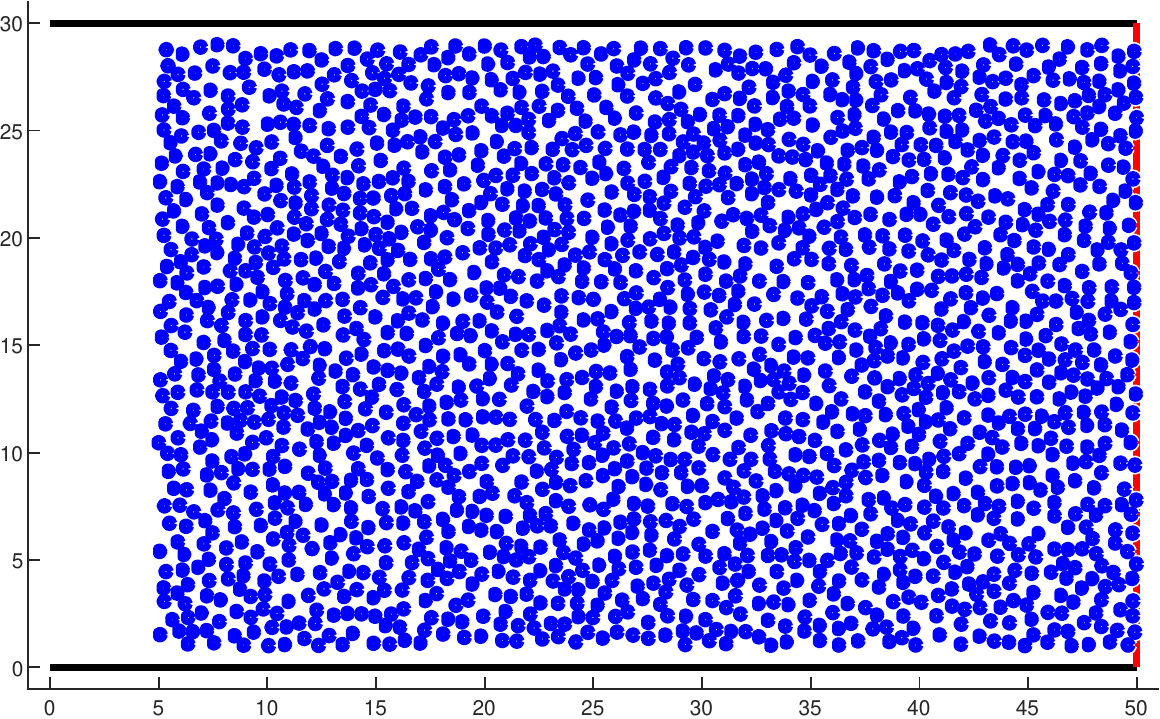} 
\includegraphics[width=0.49\textwidth]{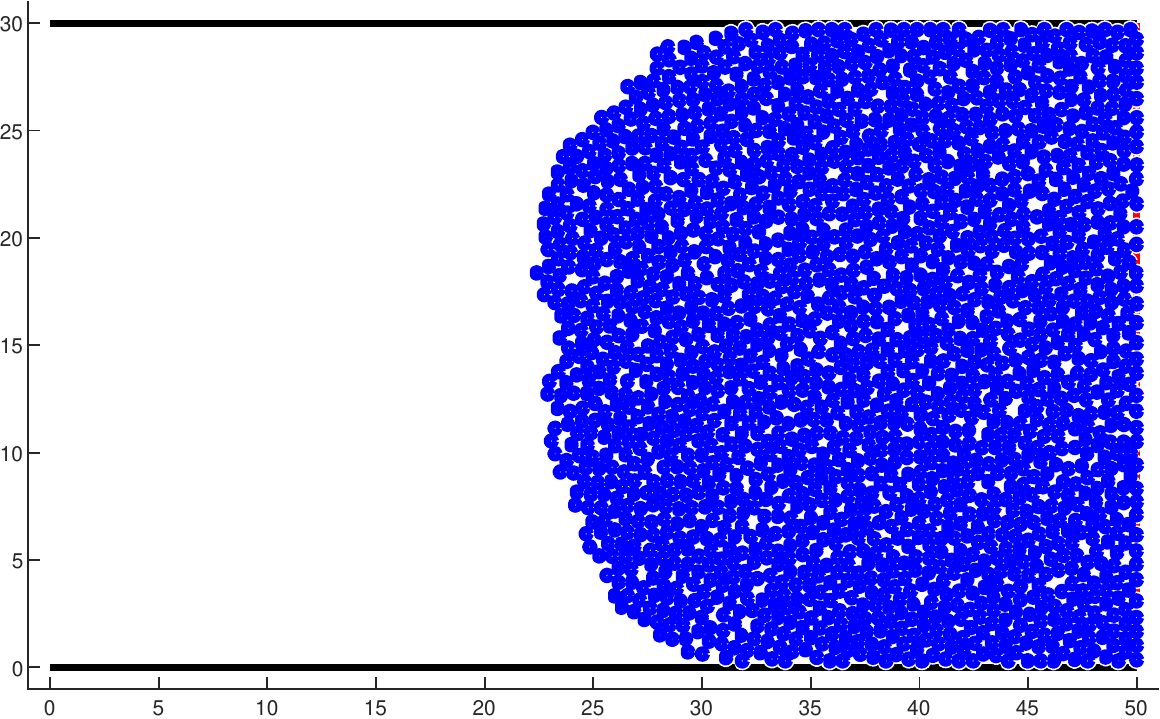} 
\includegraphics[width=0.49\textwidth]{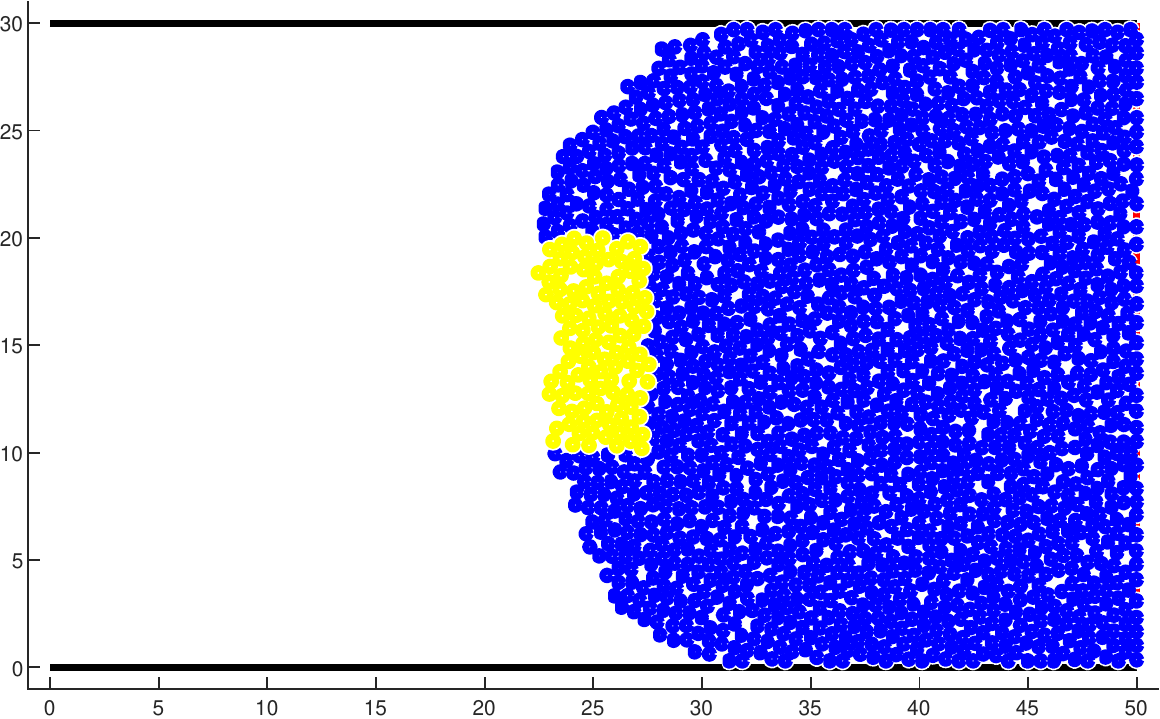} 
\includegraphics[width=0.49\textwidth]{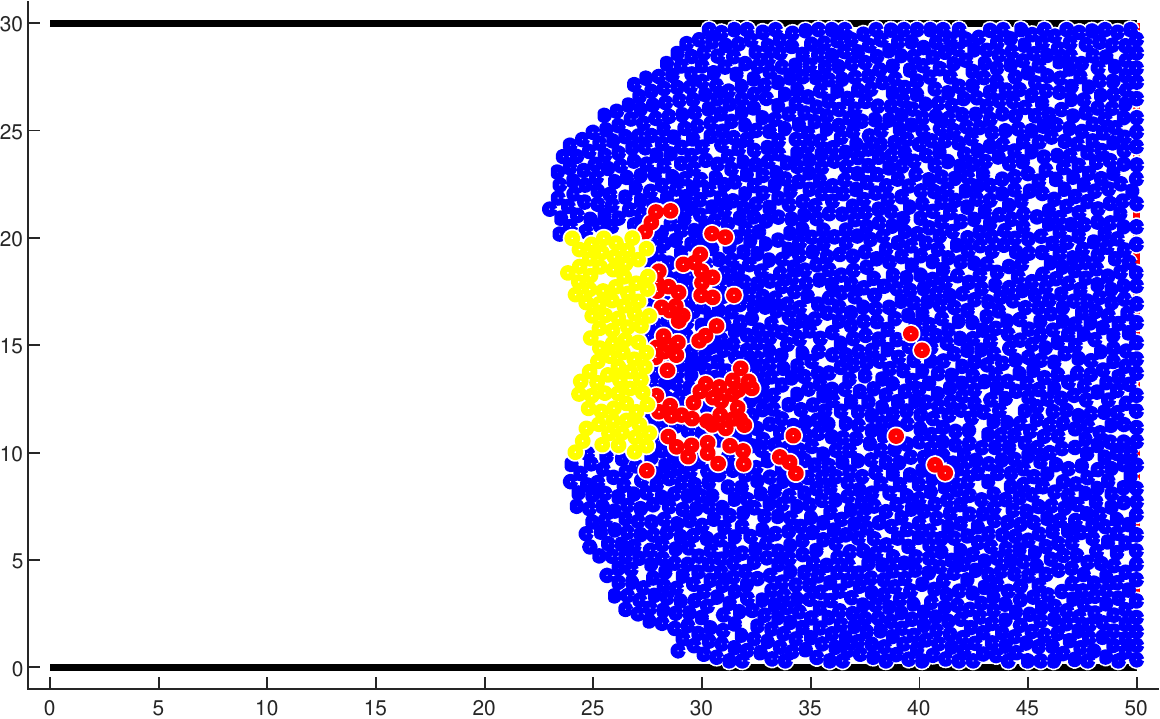} 
\includegraphics[width=0.49\textwidth]{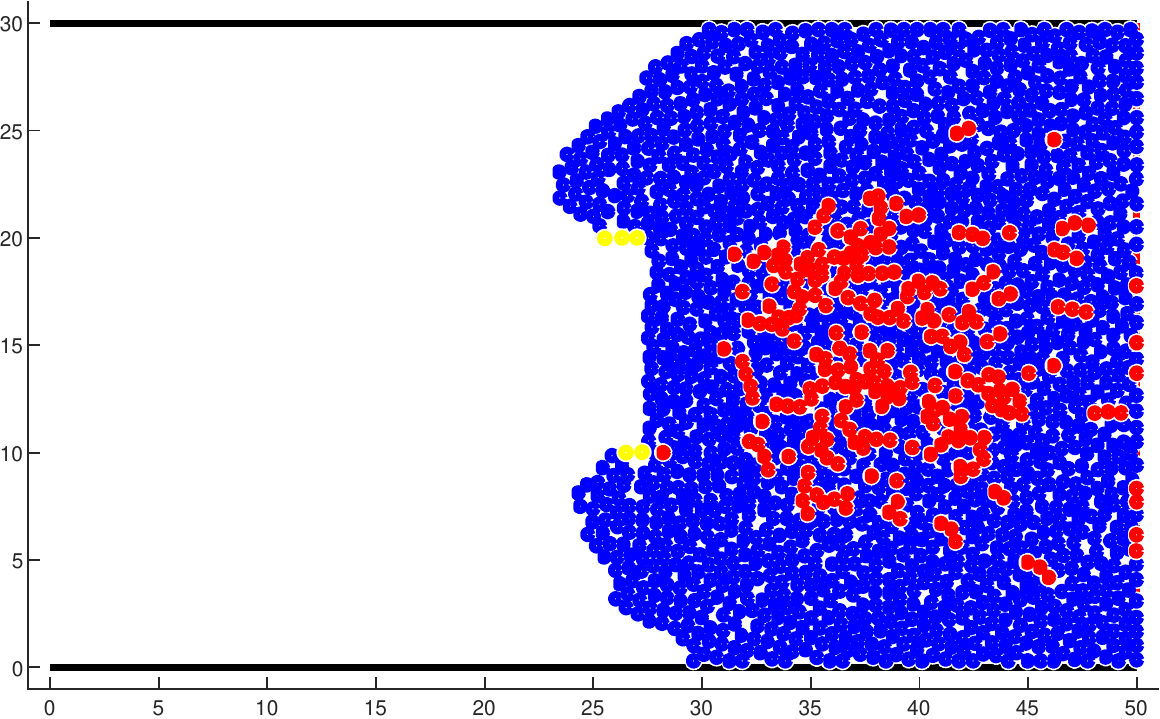} 
\includegraphics[width=0.49\textwidth]{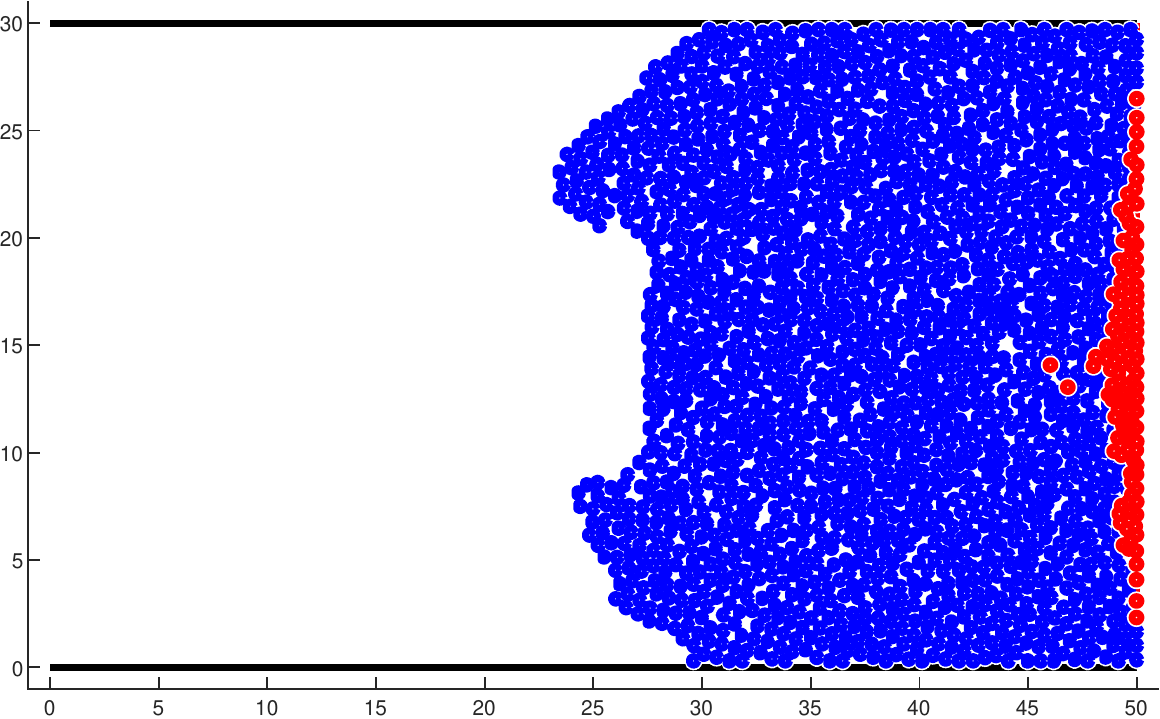} 
\caption{Test 4. \textit{From left to right, top to bottom}: six snapshots of the simulation at time step $n=1, 299, 301, 320, 375, 450$. Yellow circles are the pedestrians who trigger the wave. Red circles are pushed pedestrians. The wave propagates rightward until it reaches the boundary.}
\label{fig:T3}
\end{figure}

In this test we try to reproduce at a qualitative level the `concert with shockwave' scenario simulated in \cite{vantoll2021CeG}. 
We consider again the scenario of Test 1, but this time we have 2000 pedestrians and the corridor size is 50 m $\times$ 30 m.
A group of people, located in $[0,27.5]\times [10,20]$ start pushing at time step $n=300$ and then they stop once they reach the line $x=27.5$.
Pushing behavior is initiated by imposing $\da_k=\Dminimal$.
Fig.\ \ref{fig:T3} shows the result.

This perturbation creates a material wave which propagates rightward in the crowd. Conversely to what happen in simulations in \cite{vantoll2021CeG}, the wave does not come back after hitting the right boundary.

To  better appreciate the propagation of the wave we have preferred to create a homogeneous crowd (i.e., with constant density) at equilibrium. To this end, we have modified \texttt{BLOCK 1} using $\dn$ instead of $\df$. Doing this, pedestrians move forward taking into account the group mates behind them too, and they reach an equilibrium with constant density (see the difference with final distribution in Fig.\ \ref{fig:T1a_corridor}, where rightmost people are more compressed than leftmost people).  
For the same reason, we have chosen 
$\Dcomfort=\Dcontact=0.6$, so that agents do not leave much space between them, facilitating the propagation of the wave.


\subsection{Test 5: Room evacuation}
\begin{figure}[h!]
	\centering
	\includegraphics[width=0.53\textwidth]{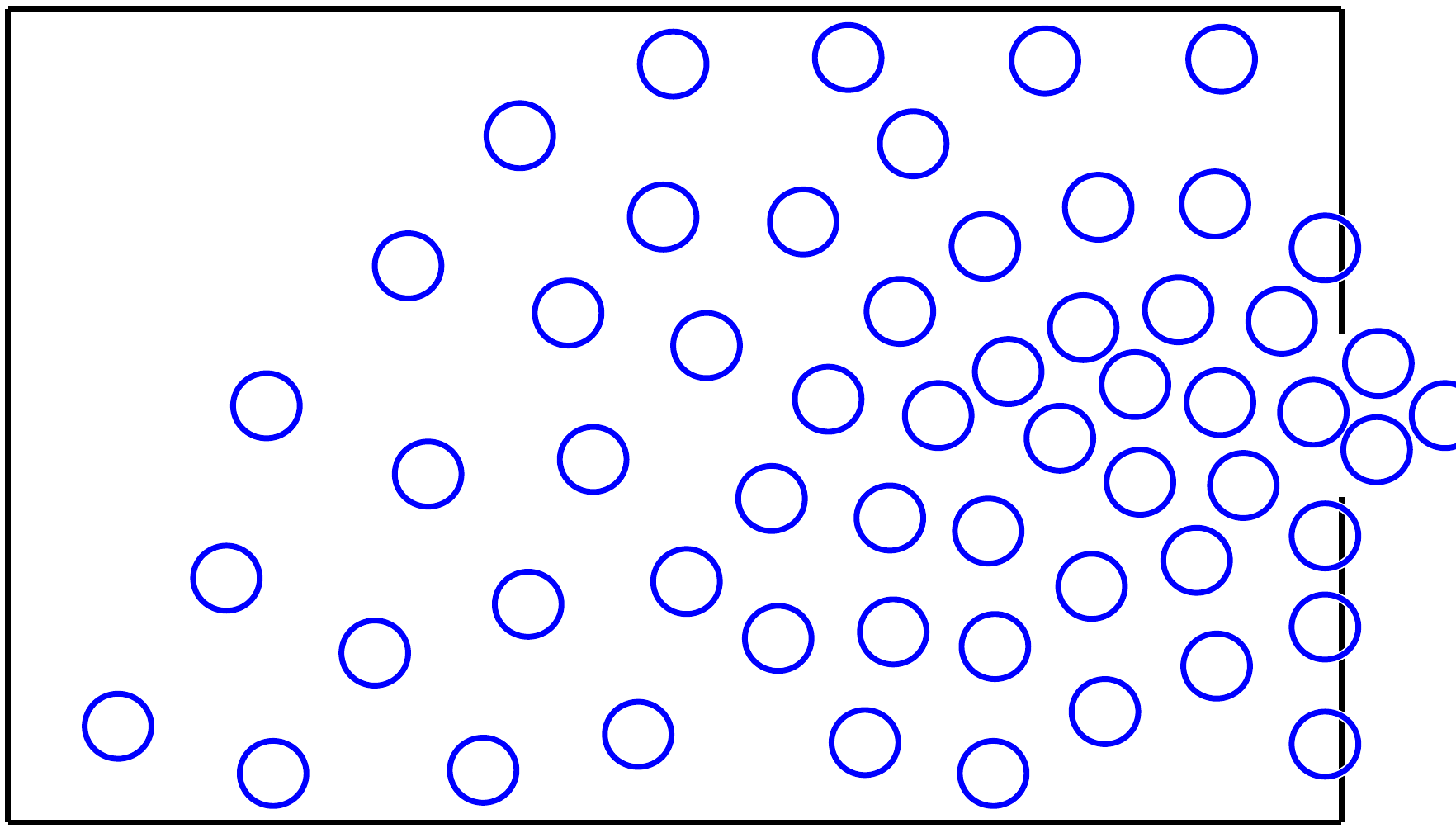} \ \ \
	\includegraphics[width=0.41\textwidth]{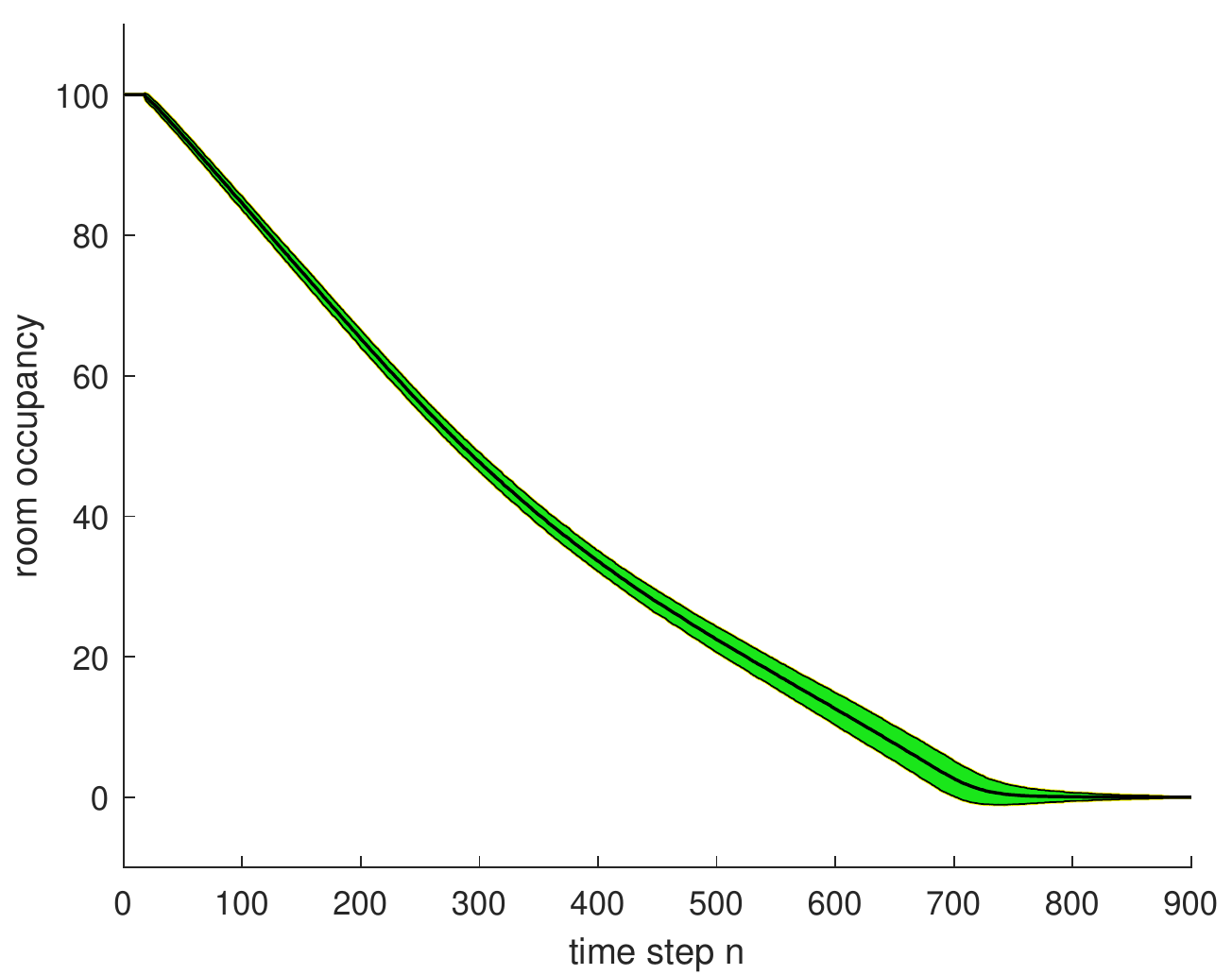} 
	\caption{Test 5.  \textit{Left}: a snapshot of the simulation.  \textit{Right}: average number of people inside the room as a function of time step $n$ (the green strip represents the area $\mu\pm\sigma$ over 300 runs).}
	\label{fig:T4iso}
\end{figure}

The flux through a door is investigated in the scenario suggested in Test 11 of ISO 20414 \cite{ISO20414}: 100 people have to leave a room of size 8 m $\times$ 5 m with a 1 m exit located centrally on the rightward 5 m wall. Here $\Sref=0.6$ m/s.
Fig.\ \ref{fig:T4iso}-left shows a snapshot of the simulation. 
Fig.\ \ref{fig:T4iso}-right shows the number of people in the room as a function of time step $n$ ($\mu\pm\sigma$ over 300 runs). 
In 75\% of runs the room is empty after 740 time steps (74 s), corresponding to an average flux of 1.35 ped/s/m.
%

%
%


\subsection{Test 6: Corner}\label{sec:T5}
\begin{figure}[h!]
	\centering
	\includegraphics[width=0.49\textwidth]{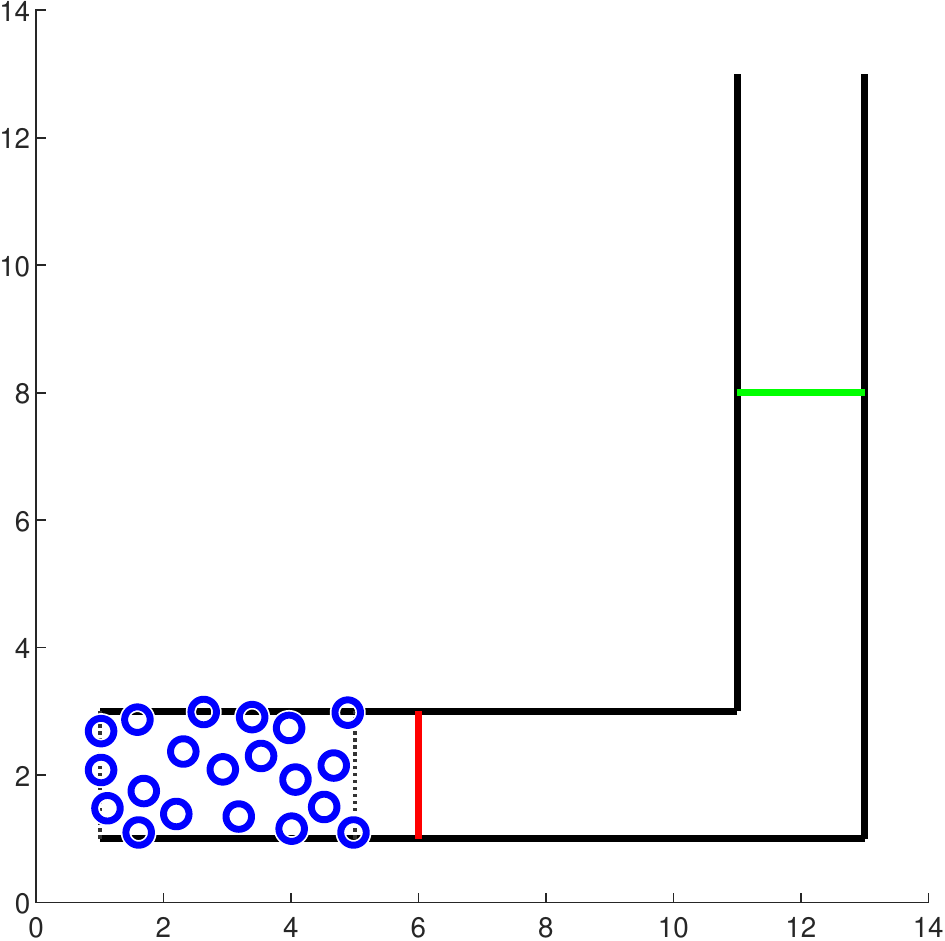} 
	\includegraphics[width=0.49\textwidth]{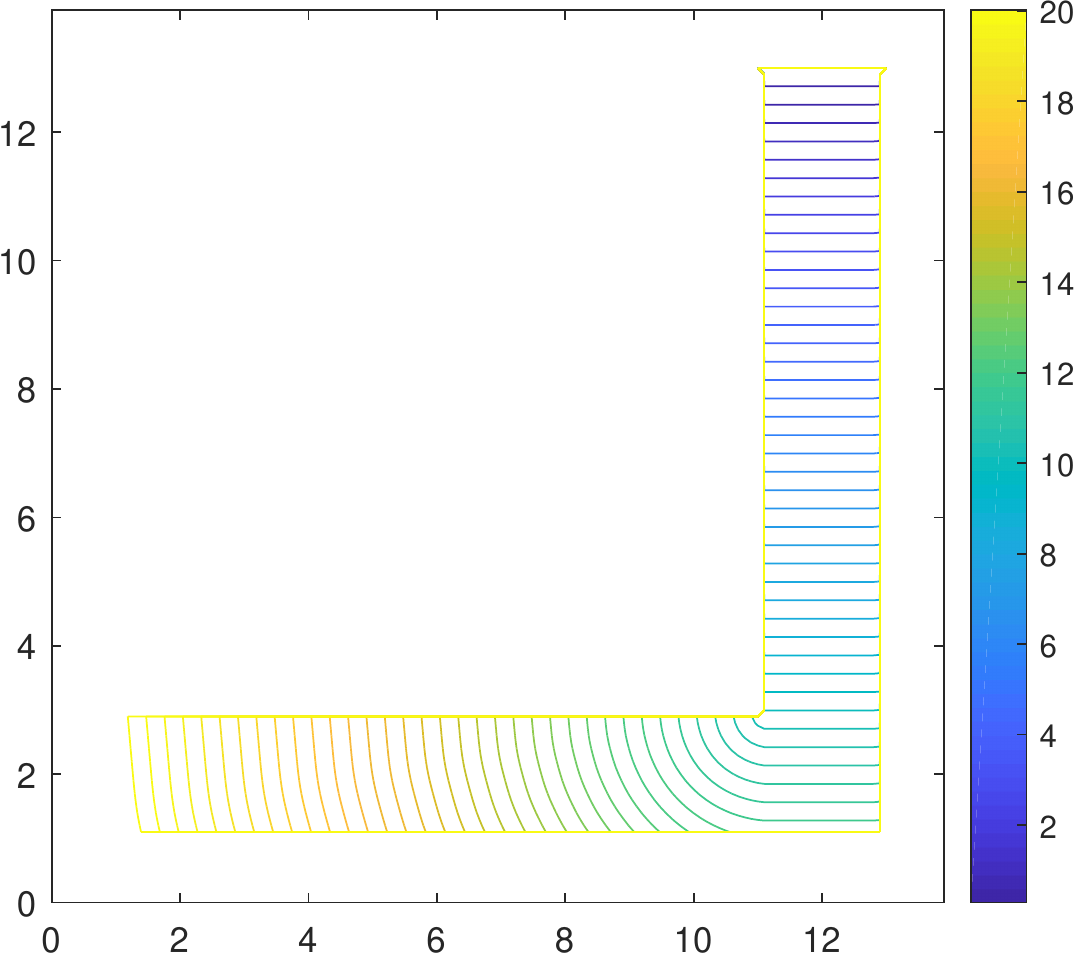} \\
	\includegraphics[width=0.49\textwidth]{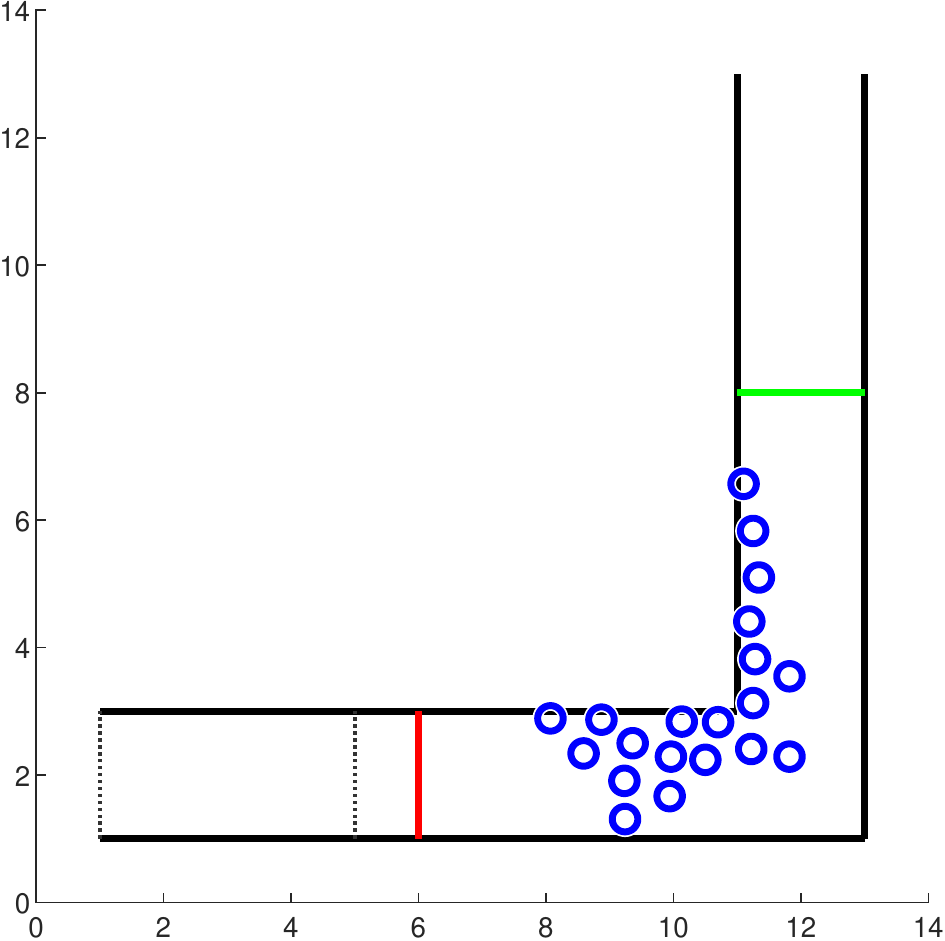} 
	\includegraphics[width=0.49\textwidth]{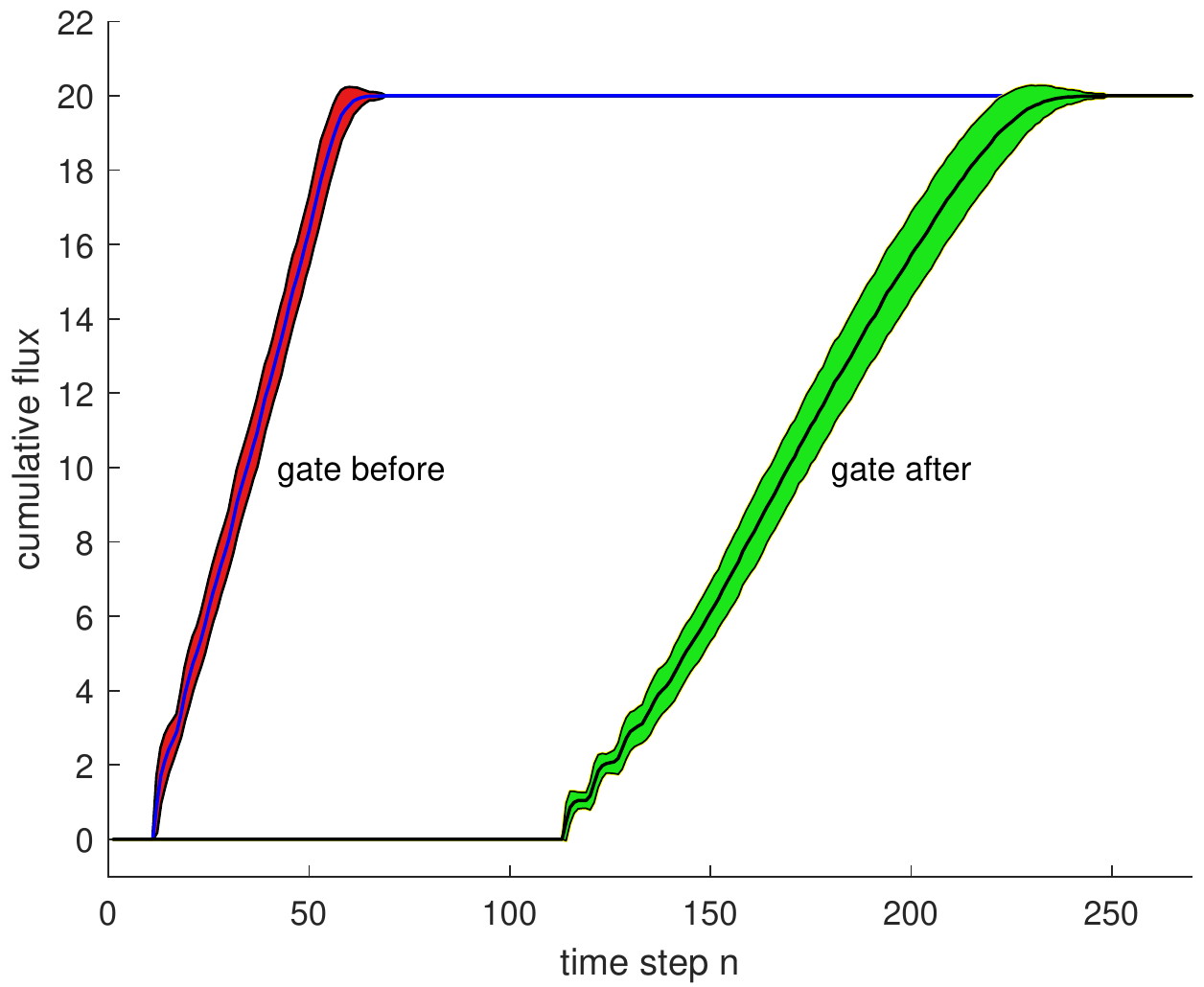} 
	\caption{Test 6. 
		\textit{Top-left}: corner scenario as in Test 4 in \cite{ISO20414}.
		\textit{Top-right}: level sets of the minimum time function, solution to the equation (\ref{ee}). Colormap units are in m/s.
		\textit{Bottom-left}: a snapshot of the simulation.
		\textit{Bottom-right}: average  cumulative number of passages across the red and the green lines. The red and the green strips represent the area $\mu\pm\sigma$ over 300 runs.
		}
	\label{fig:T5domain}
\end{figure}

This is the Test 4 in \cite{ISO20414}. The scenario is a corridor (width = 2 m, length = $[20,24]$ m) with a $90^\circ$ corner in the middle. 20 pedestrians walk through the corridor with $\Sref=1$ m/s from the bottom-left side to the the top-right side. 
They are initially located in random positions within 4 meters from the beginning, see Fig.\ \ref{fig:T5domain}(top-left). Here we have necessarily decreased $\Dcomfort$ to 0.6 m in order to let all people be ``comfortably'' confined in the starting area.

The optimal direction of motion is computed by solving the eikonal equation (\ref{ee}). 
Fig.\ \ref{fig:T5domain}(top-right) shows the level sets of the minimum time function $u$, which gives the minimal time to reach the target moving along the fastest path. 
The optimal direction $\mathbf e$ is given, at any point, by the direction orthogonal to the level sets.

Fig.\ \ref{fig:T5domain}(bottom-left) shows a snapshot of the simulation taken when pedestrians walk around the corner. The barycenter of pedestrians never enter the wall, but a small tolerance for the whole circular body exists.

Fig.\ \ref{fig:T5domain}(bottom-right) shows the cumulative number of passages across the blue line at $x_1=6$ and the green line at $x_2=8$ ($\mu\pm\sigma$ over 300 runs). The slope of the graphs corresponds to the flux through the lines. It can be seen that the corner correctly causes a bottleneck.

\begin{figure}[h!]
	\centering
	\includegraphics[width=0.49\textwidth]{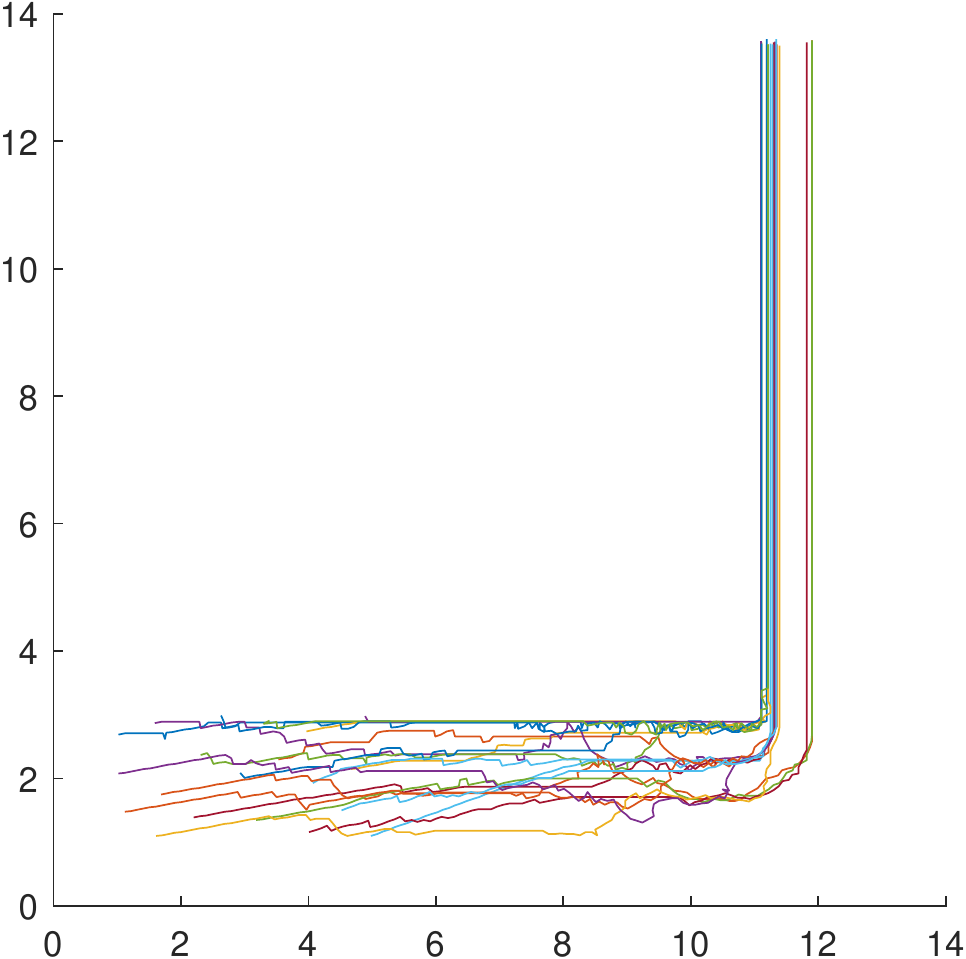} 
	\includegraphics[width=0.49\textwidth]{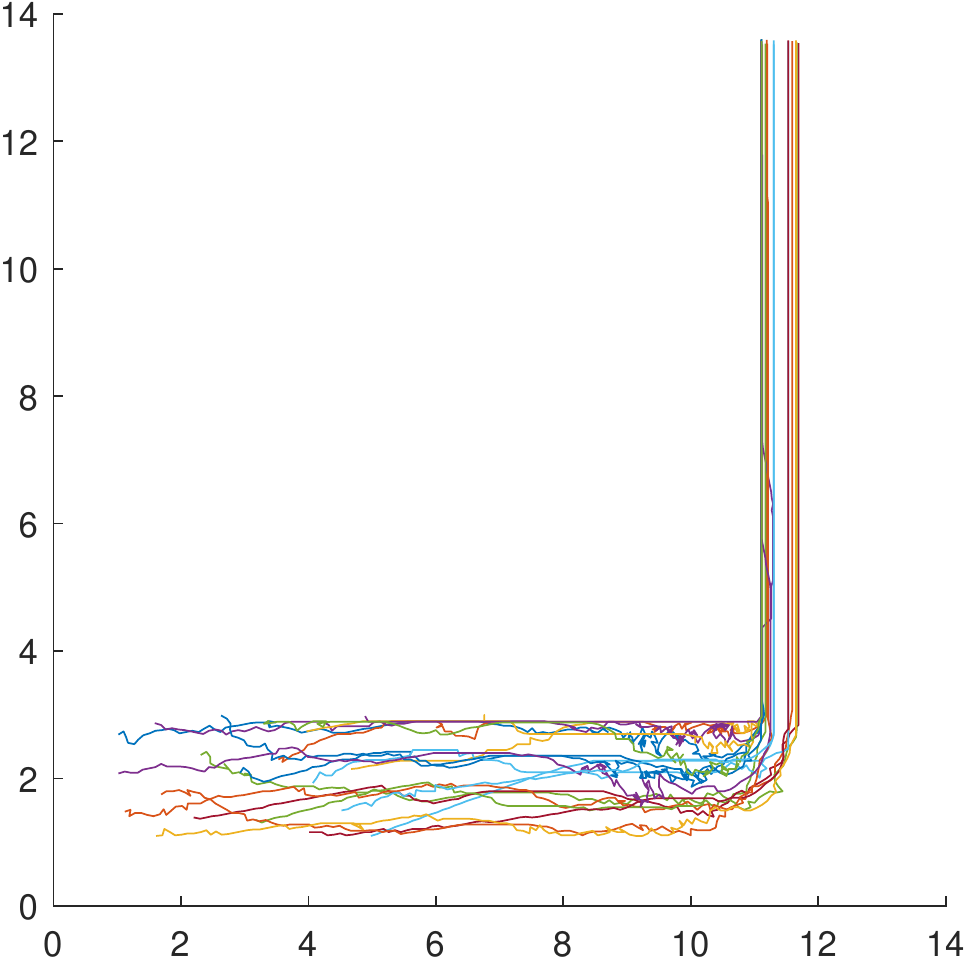} 
	\caption{Test 6. 
	\textit{Left}: trajectories of the 20 pedestrians in the case the optimal direction of motion $\mathbf e$ is computed by the equation (\ref{ee}).
	\textit{Right}: Same plot but $\mathbf e$ is computed by the equation (\ref{ee-modif}). 
}
\label{fig:T5traj}
\end{figure}

Fig.\ \ref{fig:T5traj} shows the trajectories of the 20 pedestrians in the case the optimal direction of motion $\mathbf e$ is computed by the equation (\ref{ee}) (on the left) and by the equation (\ref{ee-modif}) (on the right). 
In this second case we have removed all the points $\mathbf x$ occupied by the pedestrians (i.e.\ the circles of radius $\Dminimal$ centered at $\mathbf X_k,\ \forall k$) from the domain $\Omega$ and we have inserted them in the obstacles set $\mathcal O$.
This trick allowed us not to define the function $s(\rho)$ (and then a fundamental diagram). 
It can be seen that, in the second case, the space is better used,  especially around the corner, since some pedestrians understand that it is preferable to walk along a longer, but less crowded, path. The result can be compared with that in \cite{hartmann2014proc} where equation (\ref{ee-modif}) is also used.

%
%
%
%
%

\subsection{Test 7: Counterflow}
In the last test we simulate the classical scenario of a counterflow. 
In a corridor 60 m $\times$ 10 m two populations of pedestrians move one against the other. 
The aim is to verify that, once they meet at the center, a self-organizing lane pattern appears, i.e.\ pedestrians rearrange in parallel lanes to better exploit the shared space.

\begin{figure}[h!]
	\centering
	\includegraphics[width=0.99\textwidth]{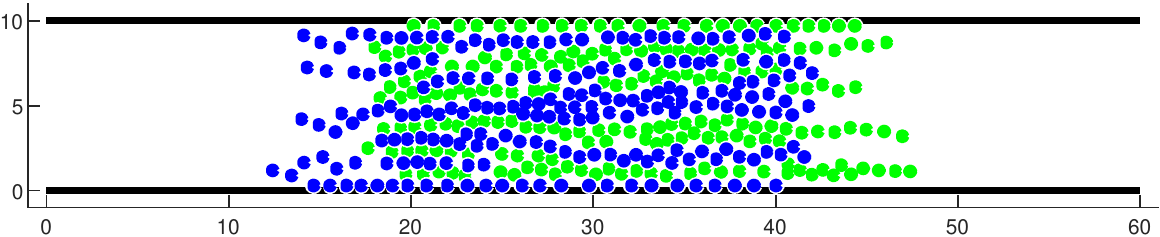} 
	\includegraphics[width=0.99\textwidth]{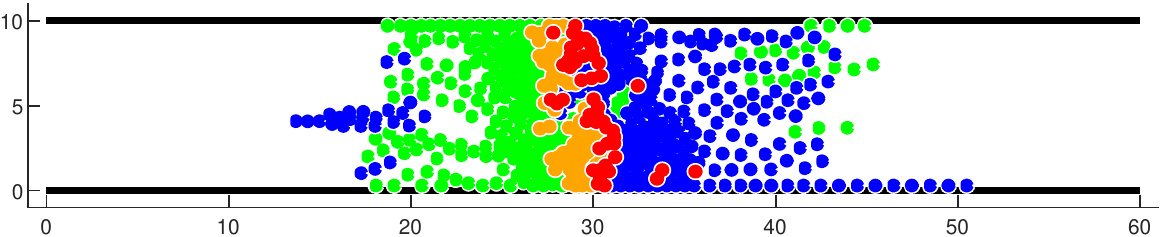} 
	\caption{Test 7. Lane formation in a corridor. 
		\textit{Top}: 200 pedestrians move leftward  (blue circles), while 200 pedestrians move rightward (green circles). 
	\textit{Bottom}: 300 pedestrians move leftward (blue circles, red if pushed), while 300 green pedestrians move rightward (green circles, orange if pushed). 
}
	\label{fig:T6}
\end{figure}

The algorithm presented in this paper strongly relies on the front/back anisotropy and it is not able, as is, to deal with people approaching from ahead. Distinguishing the two populations in the dynamics is probably the better way to proceed, in this way people can react differently to mates belonging to their own population or to the other one. Alternatively, we propose here a tiny modification of the algorithm that can serve as a workaround: in the first step of the flowchart (Fig.\ \ref{fig:diagrammadiflusso}) the accepted distance $\da_k^{n+1}$ takes the value of $\dn_k$ rather than $\db_k$, regardless of the condition involving $\alpha$. Doing this, the algorithm becomes symmetric w.r.t.\ front and back. 
We performed the test with $N=2\times200$ and $N=2\times300$ agents. 
Results are shown in Fig.\ \ref{fig:T6}.

In the first case, the two populations are able to reach their respective targets and parallel lanes show up.
In the second case, instead, only some agents belonging to the two populations are able to find a way to pass; after that, a blockage in the center appears.

%
%
%
%
%
\section{Limitations} The model proposed in this paper is presented in a very basic form in order to better highlight its features. At present stage, it has some limitations which can be overcome with additional features, at the cost of increasing complexity:
	\begin{itemize}
		\item Pedestrians have circular shape. Although we preserve the body orientation (front/back), we are not able to properly describe anisotropic torso dynamics and body rotation.
		\item Pushing forces are limited to the direction joining the pushing and the pushed agents. Pushing forces cannot act in lateral direction (for example between agents' shoulders). 
		\item Agents positions are updated one after the others, in a fixed order. 
		\item The model does not guarantee that agents do not block each other for indefinite time. This happens when two or more agents cannot move toward the target without getting too close to other agents. This fact can be sometimes observed in the scenarios presented in Test 5, near the door, and in Test 6, at corner. To fix this one should introduce symmetry-breaking priorities among agents.
		\item Pedestrians do not have predictive capabilities and do not forecast possible collisions with others. Ideas from, e.g., the recent paper \cite{xu2021TRC} could be borrowed here.
		\item As it is clear from the last Test 7, pedestrians tend to queue up along the boundaries in a nonrealistic manner. Some symmetry-breaking features should be introduced near walls.
		\item The model is particularly suited for noncooperative situations, in which pedestrians try to keep their position in a queue or in a flow of people. This is clearly not the case during leisure walks or other relaxed situations in which being overtaken is not an issue \textit{per se}.
	\end{itemize}

\section{Conclusions}
In this paper we have introduced a distance-based zeroth-order discrete-in-time model which is able to reproduce, in a microscopic setting, some classical self-organizing patterns typically observed in crowds, 
as well as the propagation of material waves due to pushing behavior.
 
 As the velocity-based and optimal step models, the proposed model avoids classical artifacts of the social force model like inertia and oscillatory movements. 
 Moreover, it is easier to calibrate because its parameters are more directly measurable (being physical distances and not imaginary forces).  
 This makes easier to meet verification \& validation protocols such the one contained in the recent International Standard ISO 20414 \cite{ISO20414}.
 
Remarkably, although the model is based on very basic heuristics, it is also able to reproduce the concave/concave fundamental diagram, showing a good ability in catching the transition free $\rightarrow$ congested $\rightarrow$ highly congested scenarios.

The model also allows to describe the fact that persons in the same scenario, as well as the same person in different scenarios, can have different degrees of tolerance regarding the interpersonal distances.
In addition, parameters can change as a response to external stimuli and mutated priorities.


In conclusions, we think that this study makes a step forward the needs of practitioners involved in low and high density crowd management, since they require easy calibration and, even more, a strong controllability of the simulation through its parameters, and a good scalability with the number of agents and domain size.

\section{Authors' contribution}
G.A.\ and A.M.\ proposed the research topic, suggested the numerical tests, interpreted the results, and proofread the paper.
E.C.\ and M.M.\ developed the model, wrote the numerical code, performed numerical tests, interpreted the results, and wrote the paper.

\section*{Funding}
This work was carried out within the research project ``SMARTOUR: Intelligent Platform for Tourism'' (No.\ SCN\_00166) funded by the Ministry of University and Research with the Regional Development Fund of European Union (PON Research and Competitiveness 2007--2013). 

E.C. would also like to thank the Italian Ministry of Instruction, University and Research (MIUR) to support this research with funds coming from PRIN Project 2017 (No.\ 2017KKJP4X entitled ``Innovative numerical methods for evolutionary partial differential equations and applications'').

E.C.\ and M.M.\ are members of the INdAM Research group GNCS.

\bibliographystyle{elsarticle-num}
\bibliography{biblio}

\begin{thebibliography}{10}
\expandafter\ifx\csname url\endcsname\relax
  \def\url#1{\texttt{#1}}\fi
\expandafter\ifx\csname urlprefix\endcsname\relax\def\urlprefix{URL }\fi
\expandafter\ifx\csname href\endcsname\relax
  \def\href#1#2{#2} \def\path#1{#1}\fi

\bibitem{reportUK}
{Variuos authors}, Understanding crowd behaviour, Tech. rep., University of
  Leeds, UK.
  https://www.gov.uk/government/publications/understanding-crowd-behaviours-documents
  (2009).

\bibitem{hirai1975proc}
K.~Hirai, K.~Tarui, A simulation of the behavior of a crowd in panic, in:
  Proceedings of the 1975 Int.\ Conf.\ on Cybernetics and Society, 1975, pp.
  409--411.

\bibitem{okazaki1979TAIJp1}
S.~Okazaki, A study of pedestrian movement in architectural space, part 1:
  {P}edestrian movement by the application of magnetic model, Trans.\ of A.I.J.
  283 (1979) 111--119.

\bibitem{henderson1974TR}
L.~F. Henderson, On the fluid mechanics of human crowd motion, Transpn. Res. 8
  (1974) 509--515.

\bibitem{aghamohammadi2020TRB}
R.~Aghamohammadi, J.~A. Laval, Dynamic traffic assignment using the macroscopic
  fundamental diagram: {A} {R}eview of vehicular and pedestrian flow models,
  Transportation Res. B 137 (2020) 99--118.
\newblock \href {http://dx.doi.org/10.1016/j.trb.2018.10.017}
  {\path{doi:10.1016/j.trb.2018.10.017}}.

\bibitem{bellomo2011SR}
N.~Bellomo, C.~Dogbe, On the modeling of traffic and crowds: {A} survey of
  models, speculations, and perspectives, SIAM Review 53 (2011) 409--463.

\bibitem{bellomo2022M3AS}
N.~Bellomo, L.~Gibelli, A.~Quaini, A.~Reali, Towards a mathematical theory of
  behavioral human crowds, Mathematical Models and Methods in Applied Sciences
  32 (2022) 321--358.
\newblock \href {http://dx.doi.org/10.1142/S0218202522500087}
  {\path{doi:10.1142/S0218202522500087}}.

\bibitem{chen2018TR}
X.~Chen, M.~Treiber, V.~Kanagaraj, H.~Li, Social force models for pedestrian
  traffic -- state of the art, Transport Reviews 38 (2018) 625--653.
\newblock \href {http://dx.doi.org/10.1080/01441647.2017.1396265}
  {\path{doi:10.1080/01441647.2017.1396265}}.

\bibitem{corbetta2023AR}
A.~Corbetta, F.~Toschi, Physics of human crowds, Annual Review of Condensed
  Matter Physics 14~(1) (2023) in press.
\newblock \href {http://dx.doi.org/10.1146/annurev-conmatphys-031620-100450}
  {\path{doi:10.1146/annurev-conmatphys-031620-100450}}.

\bibitem{dong2020TITS}
H.~Dong, M.~Zhou, Q.~Wang, X.~Yang, F.-Y. Wang, State-of-the-art pedestrian and
  evacuation dynamics, IEEE Transactions on Intelligent Transportation Systems
  21 (2020) 1849--1866.
\newblock \href {http://dx.doi.org/10.1109/TITS.2019.2915014}
  {\path{doi:10.1109/TITS.2019.2915014}}.

\bibitem{duives2013TRC}
D.~C. Duives, W.~Daamen, S.~P. Hoogendoorn, State-of-the-art crowd motion
  simulation models, Transportation Res. C 37 (2013) 193--209.
\newblock \href {http://dx.doi.org/10.1016/j.trc.2013.02.005}
  {\path{doi:10.1016/j.trc.2013.02.005}}.

\bibitem{eftimie2018chapter}
R.~Eftimie, Multi-dimensional transport equations, in: Hyperbolic and Kinetic
  Models for Self-organised Biological Aggregations, Springer, 2018, pp.
  153--193.

\bibitem{li2019PhA}
Y.~Li, M.~Chen, Z.~Dou, X.~Zheng, Y.~Cheng, A.~Mebarki, A review of cellular
  automata models for crowd evacuation, Physica A 526 (2019) 120752.
\newblock \href {http://dx.doi.org/10.1016/j.physa.2019.03.117}
  {\path{doi:10.1016/j.physa.2019.03.117}}.

\bibitem{martinezgil2017CS}
F.~Martinez-Gil, M.~Lozano, I.~Garc{\'\i}a-Fern{\'a}ndez, F.~Fern{\'a}ndez,
  Modeling, evaluation, and scale on artificial pedestrians: {A} literature
  review, ACM Comput. Surv. 50 (2017) 72.
\newblock \href {http://dx.doi.org/10.1145/3117808}
  {\path{doi:10.1145/3117808}}.

\bibitem{papadimitriou2009TRF}
E.~Papadimitriou, G.~Yannis, J.~Golias, A critical assessment of pedestrian
  behaviour models, Transportation Res. F 12 (2009) 242--255.

\bibitem{yang2020GM}
S.~Yang, T.~Li, X.~Gong, B.~Peng, J.~Hu, A review on crowd simulation and
  modeling, Graphical Models 111 (2020) 101081.
\newblock \href {http://dx.doi.org/10.1016/j.gmod.2020.101081}
  {\path{doi:10.1016/j.gmod.2020.101081}}.

\bibitem{haghani2020SSp1}
M.~Haghani, Empirical methods in pedestrian, crowd and evacuation dynamics:
  Part {I}. {E}xperimental methods and emerging topics, Safety Science 129
  (2020) 104743.
\newblock \href {http://dx.doi.org/10.1016/j.ssci.2020.104743}
  {\path{doi:10.1016/j.ssci.2020.104743}}.

\bibitem{haghani2020SSp2}
M.~Haghani, Empirical methods in pedestrian, crowd and evacuation dynamics:
  Part {II}. {F}ield methods and controversial topics, Safety Science 129
  (2020) 104760.
\newblock \href {http://dx.doi.org/10.1016/j.ssci.2020.104760}
  {\path{doi:10.1016/j.ssci.2020.104760}}.

\bibitem{haghani2021PhA}
M.~Haghani, The knowledge domain of crowd dynamics: {A}natomy of the field,
  pioneering studies, temporal trends, influential entities and outside-domain
  impact, Physica A 580 (2021) 126145.
\newblock \href {http://dx.doi.org/10.1016/j.physa.2021.126145}
  {\path{doi:10.1016/j.physa.2021.126145}}.

\bibitem{cristianibook}
E.~Cristiani, B.~Piccoli, A.~Tosin, Multiscale Modeling of Pedestrian Dynamics,
  {Modeling, Simulation \& Applications}, Springer, 2014.

\bibitem{rosinibook}
M.~D. Rosini, Macroscopic models for vehicular flows and crowd dynamics:
  {T}heory and applications, Springer, 2013.

\bibitem{kachroobook}
P.~Kachroo, S.~J. Al-nasur, S.~A. Wadoo, A.~Shende, Pedestrian dynamics.
  {F}eedback control of crowd evacuation, Understanding Complex Systems,
  Springer-Verlag, Berlin Heidelberg, 2008.

\bibitem{maurybook}
B.~Maury, S.~Faure, Crowds in equations. {A}n introduction to the microscopic
  modeling of crowds, World Scientific, 2019.

\bibitem{helbing2001RMP}
D.~Helbing, Traffic and related self-driven many-particle systems, Rev. Mod.
  Phys. 73 (2001) 1067--1141.

\bibitem{seitz2012PRE}
M.~J. Seitz, G.~K{\"o}ster, Natural discretization of pedestrian movement in
  continuous space, Physical Review E 86 (2012) 046108.
\newblock \href {http://dx.doi.org/10.1103/PhysRevE.86.046108}
  {\path{doi:10.1103/PhysRevE.86.046108}}.

\bibitem{dietrich2014JCS}
F.~Dietrich, G.~K{\"o}ster, M.~Seitz, I.~{von Sivers}, Bridging the gap: From
  cellular automata to differential equation models for pedestrian dynamics,
  Journal of Computational Science 5 (2014) 841--846.
\newblock \href {http://dx.doi.org/10.1016/j.jocs.2014.06.005}
  {\path{doi:10.1016/j.jocs.2014.06.005}}.

\bibitem{seitz2015PhA}
M.~J. Seitz, F.~Dietrich, G.~K{\"o}ster, The effect of stepping on pedestrian
  trajectories, Physica A 421 (2015) 594--604.
\newblock \href {http://dx.doi.org/10.1016/j.physa.2014.11.064}
  {\path{doi:10.1016/j.physa.2014.11.064}}.

\bibitem{vonsivers2015TRB}
I.~{von Sivers}, G.~K{\"o}ster, Dynamic stride length adaptation according to
  utility and personal space, Transportation Res. B 74 (2015) 104--117.
\newblock \href {http://dx.doi.org/10.1016/j.trb.2015.01.009}
  {\path{doi:10.1016/j.trb.2015.01.009}}.

\bibitem{paris2007EUROGRAPHICS}
S.~Paris, J.~Pettr{\'e}, S.~Donikian, Pedestrian reactive navigation for crowd
  simulation: {A} predictive approach, in: Computer Graphics Forum, Vol.~26,
  Wiley Online Library, 2007, pp. 665--674.

\bibitem{tang2016PhA}
M.~Tang, H.~Jia, B.~Ran, J.~Li, Analysis of the pedestrian arching at
  bottleneck based on a bypassing behavior model, Physica A 453 (2016)
  242--258.
\newblock \href {http://dx.doi.org/10.1016/j.physa.2016.02.044}
  {\path{doi:10.1016/j.physa.2016.02.044}}.

\bibitem{antonini2006TRB}
G.~Antonini, M.~Bierlaire, M.~Weber, Discrete choice models of pedestrian
  walking behavior, Transportation Res. B 40 (2006) 667--687.
\newblock \href {http://dx.doi.org/10.1016/j.trb.2005.09.006}
  {\path{doi:10.1016/j.trb.2005.09.006}}.

\bibitem{robin2009TRB}
T.~Robin, G.~Antonini, M.~Bierlaire, J.~Cruz, Specification, estimation and
  validation of a pedestrian walking behavior model, Transportation Res. B 43
  (2009) 36--56.
\newblock \href {http://dx.doi.org/10.1016/j.trb.2008.06.010}
  {\path{doi:10.1016/j.trb.2008.06.010}}.

\bibitem{seitz2016JRSI}
M.~J. Seitz, N.~W.~F. Bode, G.~K{\"o}ster, How cognitive heuristics can explain
  social interactions in spatial movement, J. R. Soc. Interface 13 (2016)
  20160439.
\newblock \href {http://dx.doi.org/10.1098/rsif.2016.0439}
  {\path{doi:10.1098/rsif.2016.0439}}.

\bibitem{haghani2019JAT}
M.~Haghani, E.~Cristiani, N.~W.~F. Bode, M.~Boltes, A.~Corbetta, Panic,
  irrationality, and herding: {T}hree ambiguous terms in crowd dynamics
  research, Journal of Advanced Transportation 2019 (2019) 9267643.
\newblock \href {http://dx.doi.org/10.1155/2019/9267643}
  {\path{doi:10.1155/2019/9267643}}.

\bibitem{jin2021JAT}
C.-J. Jin, X.~Shi, T.~Hui, D.~Li, K.~Ma, The automatic detection of pedestrians
  under the high-density conditions by deep learning techniques, Journal of
  Advanced Transportation 2021 (2021) 1396326.
\newblock \href {http://dx.doi.org/10.1155/2021/1396326}
  {\path{doi:10.1155/2021/1396326}}.

\bibitem{helbing2007PRE}
D.~Helbing, A.~Johansson, H.~Z. Al-Abideen, Dynamics of crowd disasters: {An}
  empirical study, Physical Review E 75 (2007) 046109.
\newblock \href {http://dx.doi.org/10.1103/PhysRevE.75.046109}
  {\path{doi:10.1103/PhysRevE.75.046109}}.

\bibitem{johansson2008ACS}
A.~Johansson, D.~Helbing, H.~Z. Al-Abideen, S.~Al-Bosta, From crowd dynamics to
  crowd safety: {A} video-based analysis, Advances in Complex Systems 11 (2008)
  497--527.
\newblock \href {http://dx.doi.org/10.1142/S0219525908001854}
  {\path{doi:10.1142/S0219525908001854}}.

\bibitem{jin2019TRC}
C.-J. Jin, R.~Jiang, S.~C. Wong, S.~Xie, D.~Li, N.~Guo, W.~Wang, Observational
  characteristics of pedestrian flows under high-density conditions based on
  controlled experiments, Transportation Res. C 109 (2019) 137--154.

\bibitem{lohner2017CD}
R.~L\"ohner, B.~Muhamad, P.~Dambalmath, E.~Haug, Fundamental diagrams for
  specific very high density crowds, Collective Dynamics 2 (2017) 1--15.
\newblock \href {http://dx.doi.org/10.17815/CD.2017.13}
  {\path{doi:10.17815/CD.2017.13}}.

\bibitem{colombo2005M2AS}
R.~M. Colombo, M.~D. Rosini, Pedestrian flows and non-classical shocks,
  Mathematical Methods in the Applied Sciences 28 (2005) 1553--1567.
\newblock \href {http://dx.doi.org/10.1002/mma.624}
  {\path{doi:10.1002/mma.624}}.

\bibitem{chalons2007SISC}
C.~Chalons, Numerical approximation of a macroscopic model of pedestrian flows,
  SIAM Journal on Scientific Computing 29 (2007) 539--555.
\newblock \href {http://dx.doi.org/10.1137/050641211}
  {\path{doi:10.1137/050641211}}.

\bibitem{kim2013SIGGRAPH}
S.~Kim, S.~J. Guy, D.~Manocha, Velocity-based modeling of physical interactions
  in multi-agent simulations, in: Proceedings of the 12th ACM
  SIGGRAPH/Eurographics symposium on computer animation, 2013, pp. 125--133.

\bibitem{kim2015TVC}
S.~Kim, S.~J. Guy, K.~Hillesland, B.~Zafar, A.~Gutub, D.~Manocha,
  Velocity-based modeling of physical interactions in dense crowds, The Visual
  Computer 31 (2015) 541--555.

\bibitem{alrashed2020CD}
M.~M. Alrashed, J.~S. Shamma, Agent based modelling and simulation of
  pedestrian crowds in panic situations, Collective Dynamics 5 (2020) 463--466.

\bibitem{helbing2000N}
D.~Helbing, I.~Farkas, T.~Vicsek, Simulating dynamical features of escape
  panic, Nature 407~(6803) (2000) 487--490.

\bibitem{helbing2002PED}
D.~Helbing, I.~J. Farkas, P.~Moln\'ar, T.~Vicsek, Simulation of pedestrian
  crowds in normal and evacuation situations, Pedestrian and Evacuation
  Dynamics 21 (2002) 21--58.

\bibitem{moussaid2011PNAS}
M.~Moussa{\"\i}d, D.~Helbing, G.~Theraulaz, How simple rules determine
  pedestrian behavior and crowd disasters, Proceedings of the National Academy
  of Sciences 108 (2011) 6884--6888.
\newblock \href {http://dx.doi.org/10.1073/pnas.1016507108}
  {\path{doi:10.1073/pnas.1016507108}}.

\bibitem{yu2007PRE}
W.~Yu, A.~Johansson, Modeling crowd turbulence by many-particle simulations,
  Physical Review E 76 (2007) 046105.
\newblock \href {http://dx.doi.org/10.1103/PhysRevE.76.046105}
  {\path{doi:10.1103/PhysRevE.76.046105}}.

\bibitem{liang2021TRB}
H.~Liang, J.~Du, S.~C. Wong, A continuum model for pedestrian flow with
  explicit consideration of crowd force and panic effects, Transportation Res.
  B 149 (2021) 100--117.
\newblock \href {http://dx.doi.org/10.1016/j.trb.2021.05.006}
  {\path{doi:10.1016/j.trb.2021.05.006}}.

\bibitem{narain2009SIGGRAPH}
R.~Narain, A.~Golas, S.~Curtis, M.~C. Lin, Aggregate dynamics for dense crowd
  simulation, in: ACM SIGGRAPH Asia 2009 papers, Association for Computing
  Machinery, New York, NY, USA, 2009, pp. 1--8.
\newblock \href {http://dx.doi.org/10.1145/1661412.1618468}
  {\path{doi:10.1145/1661412.1618468}}.

\bibitem{vantoll2021CeG}
W.~{van Toll}, T.~Chatagnon, C.~Braga, B.~Solenthaler, J.~Pettr{\'e}, {SPH}
  crowds: {A}gent-based crowd simulation up to extreme densities using fluid
  dynamics, Computers \& Graphics 98 (2021) 306--321.
\newblock \href {http://dx.doi.org/10.1016/j.cag.2021.06.005}
  {\path{doi:10.1016/j.cag.2021.06.005}}.

\bibitem{ISO20414}
{ISO 20414} {F}ire safety engineering -- {V}erification and validation protocol
  for building fire evacuation models, ISO (2020).

\bibitem{seitz2017RGP}
M.~J. Seitz, A.~Templeton, J.~Drury, G.~K{\"o}ster, A.~Philippides, Parsimony
  versus reductionism: {H}ow can crowd psychology be introduced into computer
  simulation?, Review of General Psychology 21 (2017) 95--102.

\bibitem{wang2018JSM}
C.~Wang, W.~Weng, Study on the collision dynamics and the transmission pattern
  between pedestrians along the queue, Journal of Statistical Mechanics: Theory
  and Experiment 2018 (2018) 073406.
\newblock \href {http://dx.doi.org/10.1088/1742-5468/aace27}
  {\path{doi:10.1088/1742-5468/aace27}}.

\bibitem{song2019IEEEA}
J.~Song, F.~Chen, Y.~Zhu, N.~Zhang, W.~Liu, K.~Du, Experiment calibrated
  simulation modeling of crowding forces in high density crowd, IEEE Access 7
  (2019) 100162--100173.
\newblock \href {http://dx.doi.org/10.1109/ACCESS.2019.2930104}
  {\path{doi:10.1109/ACCESS.2019.2930104}}.

\bibitem{cacace2014SISC}
S.~Cacace, E.~Cristiani, M.~Falcone, Can local single-pass methods solve any
  stationary {H}amilton--{J}acobi--{B}ellman equation?, SIAM Journal on
  Scientific Computing 36~(2) (2014) A570--A587.
\newblock \href {http://dx.doi.org/10.1137/130907707}
  {\path{doi:10.1137/130907707}}.

\bibitem{falconebook}
M.~Falcone, R.~Ferretti, Semi-{L}agrangian approximation schemes for linear and
  {H}amilton--{J}acobi equations, SIAM, 2013.

\bibitem{sethianbook}
J.~A. Sethian, Level set methods and {F}ast {M}arching methods: evolving
  interfaces in computational geometry, fluid mechanics, computer vision, and
  materials science, Cambridge University Press, 1999.

\bibitem{hughes2002TRB}
R.~L. Hughes, A continuum theory for the flow of pedestrians, Transportation
  Res. B 36~(6) (2002) 507--535.

\bibitem{hartmann2014proc}
D.~Hartmann, J.~Mille, A.~Pfaffinger, C.~Royer, Dynamic medium scale navigation
  using dynamic floor fields, in: U.~Weidmann, U.~Kirsch, M.~Schreckenberg
  (Eds.), Pedestrian and Evacuation Dynamics 2012, Springer, Cham, 2014, pp.
  1237--1249.
\newblock \href {http://dx.doi.org/10.1007/978-3-319-02447-9}
  {\path{doi:10.1007/978-3-319-02447-9}}.

\bibitem{lachapelle2011TRB}
A.~Lachapelle, M.-T. Wolfram, On a mean field game approach modeling congestion
  and aversion in pedestrian crowds, Transportation Res. B 45 (2011)
  1572--1589.
\newblock \href {http://dx.doi.org/10.1016/j.trb.2011.07.011}
  {\path{doi:10.1016/j.trb.2011.07.011}}.

\bibitem{cristiani2023CMS}
E.~Cristiani, A.~De~Santo, M.~Menci, A generalized mean-field game model for
  the dynamics of pedestrians with limited predictive abilities, Commun. Math.
  Sci. 21~(1) (2023) 65--82.

\bibitem{xu2021TRC}
Q.~Xu, M.~Chraibi, A.~Seyfried, Anticipation in a velocity-based model for
  pedestrian dynamics, Transportation Res. C 133 (2021) 103464.

\end{thebibliography}

\end{document}